\documentclass[showpacs,preprintnumbers,amssymb]{revtex4-2}

\usepackage{graphicx}
\usepackage{dcolumn}
\usepackage[dvipsnames]{xcolor}
\usepackage{bm}
\usepackage{amsmath}
\usepackage{amssymb}
\usepackage{epsfig}
\usepackage{amsfonts}
\usepackage{lineno,hyperref}
\usepackage{epsfig}
\usepackage{array}
\usepackage{float}
\usepackage{microtype}
\usepackage{subcaption}
\usepackage{multirow}
\usepackage{adjustbox}
\usepackage[english]{babel}
\usepackage{epstopdf}
\usepackage{cuted}
\usepackage{duckuments}

\usepackage{blindtext}
\usepackage{subcaption}
\usepackage[a4paper, total={6.5in, 10in}]{geometry}

\def \a{\alpha}
\def \b{\beta}
\def \l{\lambda}

\def \e{\epsilon}

\def \g{\bar{g}}

\def \k{\kappa}
\def \o{\omega}
\def \O{\Omega}

\def \p{\partial}
\def \t{\theta}

\def \nb{\nabla}

\def \be{\begin{equation}}
\def \ee{\end{equation}}
\def \ben{\begin{eqnarray}}
\def \een{\end{eqnarray}}

\def \K{\dot{\phi}^2}
\def \R{\bar{R}}

\def \half{\frac{1}{2}}
\def \sg{\sigma}

\def \T{\bar{T}}
\def \Th{\bar{\Theta}}

\def \nn{\nonumber}

\DeclareUnicodeCharacter{2212}{-}

\begin{document}

\title{Exploring Cosmological Implications of the Modified Raychaudhuri Equation in Non-Gravitating Vacuum Energy Theory}

\author{Arijit Panda}
\email{arijitpanda260195@gmail.com}
\affiliation{Department of Physics, Raiganj University, Raiganj, Uttar Dinajpur, West Bengal, India, 733 134. $\&$\\
Department of Physics, Prabhat Kumar College, Contai, Purba Medinipur, India, 721 404.}

\author{Eduardo Guendelman}
\email{guendel@bgu.ac.il}
\affiliation{Department of Physics, Ben-Gurion University of the Negev, Beer-Sheva, Israel $\&$\\
 Frankfurt Institute for Advanced Studies (FIAS),
 Ruth-Moufang-Strasse 1, 60438 Frankfurt am Main, Germany. $\&$\\
Bahamas Advanced Study Institute and Conferences, 4A Ocean Heights,
 Hill View Circle, Stella Maris, Long Island, The Bahamas.}

\author{Goutam Manna$^a$}
\email{goutammanna.pkc@gmail.com}
\altaffiliation{$^a$Corresponding author}
\affiliation{Department of Physics, Prabhat Kumar College, Contai, Purba Medinipur 721404, India $\&$\\ Institute of Astronomy Space and Earth Science, Kolkata 700054, India}

\begin{abstract}
This article investigates the modified Raychaudhuri Equation (RE) in the context of Non-Gravitating Vacuum Energy (NGVE) theory and its implications for various cosmological characteristics. The equation is formulated based on the NGVE framework, in which global scale invariance generates a unique geometry. The newly developed geometry introduces a metric that is conformally connected to the conventional metric, with the conformal factor dependent on scalar field potentials. The cosmological study is carried out under the framework of a flat Friedmann-Lemaître-Robertson-Walker (FLRW) universe. Assuming matter behaves as an ideal fluid in the modified geometry, we formulate models for conditional expansion, collapse, and steady state, governed by the scalar field ($\phi$). In this context, the caustic solution and the focusing theorem are also studied. Scalar field solutions for exponential and power-law scale factors are also derived using NGVE theory's equations of motion. Finally, graphical analysis is used to investigate the behavior of the interaction terms that appear in the modified RE under these scale factors. 
\end{abstract}

\date{\today}

\keywords{Raychaudhuri equation, NGVE theory, Dark energy, Expanding and collapsing universe }

\maketitle
\section{Introduction}
Approximately seventy years ago, General Relativity (GR) was a nascent theory and the ramifications of the Schwarzschild solution were not fully understood. The study of cosmology was nascent and the singularities present in general relativity solutions, including those in the Schwarzschild metric and cosmic models, raised significant issues. Einstein questioned whether singularities were an inescapable aspect of GR and if their existence compromised the theory's credibility.

In the early 1950s, Raychaudhuri started tackling many core difficulties in GR. In 1955, he presented a geometrical equation, now known as the Raychaudhuri equation (RE), linking Einstein's field equations to the evolution of congruences of geodesics \cite{Ray1}. The study, which was mostly driven by issues in cosmology, established the foundation for comprehending the temporal evolution of spacetime geometry. Although Raychaudhuri initially avoided assuming homogeneity or isotropy, he constructed a universe model that included time-dependent geometry, illustrating the cosmos' dynamic essence.

In 1957, he reexamined his equation and formulated it within a broader context, illustrating how the Newtonian findings of Heckmann and Schücking \cite{Heckmann1} could be extended to the completely relativistic realm \cite{Ray2}. The RE is fundamental to contemporary cosmology. It offers a mathematical framework for scrutinizing the temporal evolution of a geodesic congruence by investigating the behavior of neighbouring geodesics \cite{Ray1, Ray2, Ray3, Kar}. This equation is essential for understanding gravitational focusing and the dynamics of spacetime geometry.

In other words, we can say that the RE is a crucial and multifaceted tool for analyzing the characteristics of spacetime and the dynamics of the cosmos. It serves as an essential connection between the abstract mathematical structure of GR and its physical consequences, significantly contributing to the comprehension of singularities, spacetime geometry and cosmic development. The equation is fundamentally a geometric identity, irrespective of any particular gravitational theory. However, it gains a dynamic character when applied to the study of gravitational spacetimes, providing powerful insights into their evolution and structure \cite{Chakraborty2, Chakraborty1}.

Hawking and Penrose's groundbreaking research \cite{Hawking1, Hawking2, Penrose1,Hawking3}, released a decade later, was the first to deeply acknowledge the importance of Raychaudhuri's contribution. The Raychaudhuri equation first emerged in the physics literature during this period, along with the development and validation of the singularity theorems.

An interesting facet of the RE is its relation to the focusing theorem \cite{Penrose2}. The evolution of geodesic congruences may demonstrate either focusing or non-focusing characteristics, depending upon the underlying conditions \cite{Choudhury, Das}. The convergence condition indicates that geodesic congruences will concentrate if the strong energy condition, which relates to the Ricci tensor, is satisfied \cite{Poisson}. The modifications of RE in the non-canonical setting with applications to the cosmic behaviors are found in \cite{Panda, Panda2}.\\

Our present understanding of the cosmos, however substantial in several dimensions, remains incomplete within the confines of traditional cosmological theories, including the prevalent $\Lambda$CDM framework \cite{Straumann, Carroll, Martin, Peri}. These theories insufficiently elucidate various fundamental phenomena, including the characteristics of dark matter and dark energy \cite{Ferreira}, the disparity between matter and antimatter \cite{Robson}, the cosmological constant issue \cite{Carroll, Martin}, the dimensions and configuration of the universe and the processes underlying cosmic inflation \cite{Ancho}. Moreover, persistent challenges such as the horizon problem highlight the need for more profound understanding. The most significant difficulty in contemporary basic physics is the integration of gravity with quantum mechanics into a cohesive theoretical framework \cite{Teit}. This deficiency in our understanding indicates that much effort is still required. Possible solutions to these issues include modified theories of gravity \cite{Shankar}, scalar-tensor theories \cite{Quiros} and innovative frameworks such as non-gravitating vacuum energy (NGVE) theories \cite{Guendelman0,Guendelman1,Guendelman2,Guendelman3}, etc., each offering alternative perspectives on the fundamental structure and evolution of the cosmos.\\

This study examines NGVE theory by modifying the RE with NGVE principles to investigate many cosmological scenarios and tackle associated challenges. The NGVE theory \cite{Guendelman0,Guendelman1,Guendelman2,Guendelman3} offers a novel solution to the longstanding issue of the cosmological constant problem in physics. NGVE provides a paradigm that may harmonize quantum field theory and cosmology by modifying the interaction between vacuum energy and gravity. Although in its nascent phase, NGVE theory signifies a potential avenue for further study. Its implications include basic physics, ranging from quantum gravity to cosmic evolution, making it a compelling domain for both theoretical and observational exploration. The assumption of invariant measure $\sqrt{-g}d^D x$ in Einstein-Hilbert (E-H) action was replaced by a suitable, generally coordinated, invariant measure selection. Unlike the Einstein-Hilbert action, where a change in the Lagrangian density results in a cosmological term, this theory will not be affected by the inclusion of a constant.\\

Our goal in this study is to create a modified RE within the framework of NGVE theory and then utilize it to examine the cosmological consequences of this modified RE while keeping the cosmological principle in mind, namely a homogeneous and isotropic universe. This paper is organized as follows: In Section II we produce a brief review of NGVE theory. Under this section, subsection II-A is devoted to formulating a new (modified) geometry under this NGVE theory. In Section III we construct the Raychaudhuri equation in this newly formed geometry. Section IV describes the cosmological analysis corresponding to the modified Raychaudhuri equation. In subsection IV-A, the analysis of $f(\chi)\neq 0$ has been done. Under this subsection, we study the solution of the equation of motion and study the behavior of $f(\chi)$ for two special cases of scale factors: exponential and power law scale factors. In Subsection IV-B we study the $f(\chi)=0$ case. In Subsection IV-C, the focusing theorem is discussed, with the mention of caustic formation. Section V is the conclusion of our work.

\section{The NGVE theory}
The gravitational theories with covariant Lagrangian have the form \cite{Schutz, Weinberg, Blau, Poisson}
\ben
S_{1}=\int L\sqrt{-g}~d^{4}x
\label{1}
\een
where $g=\det g_{\mu\nu}$ and $L$ is the scalar Lagrangian which incorporates the gravitational effects. NGVE theory \cite{Guendelman0,Guendelman1,Guendelman2,Guendelman3,Guendelman5,Guendelman6,Guendelman7} allows us to use a measure field $(\Phi)$ which has similar transformation properties of $\sqrt{-g}$ and gives a new covariant formulation. The form of the measure field is given by \cite{Guendelman0,Guendelman3}
\ben
\Phi=\e^{\mu\nu\a\b}\e_{abcd}\p_{\mu}\varphi_{a}\p_{\nu}\varphi_{b}\p_{\a}\varphi_{c}\p_{\b}\varphi_{d}
\label{2}
\een
where $\varphi_{a}~(a=1,2,3,4)$ are four measure scalars. This consideration alters the action (\ref{1}) as 
\ben
S_{2}=\int L \Phi~d^{4}x
\label{3}
\een
The above formulation (Eq. \ref{3}) is assumed as the first-order consideration because the measure and the connection are independent degrees of freedom, i.e., independent of metric and matter fields. The ability to express $\Phi$ as a total derivative in the form \cite{Guendelman3}
\ben
\Phi=\p_{\mu}(\e^{\mu\nu\a\b}\e_{abcd}\varphi_{a}\p_{\nu}\varphi_{b}\p_{\a}\varphi_{c}\p_{\b}\varphi_{d})
\label{4}
\een
allows us to add a constant to the Lagrangian ($L\rightarrow L+constant$) without changing the equation of motion of the theory excluding the notion of cosmological constant. This property is well known as the shift symmetry. When we apply the shifting process of Lagrangian it gives rise to the cosmological constant in the case of the E-H action (\ref{1}). NGVE theory is free of this problem due to the shift symmetry and does not account for any extra term in the equation of motion \cite{Guendelman3}, hence the name NGVE is justified. 

Considering both the Lagrangian of Eq. (\ref{1}) and Eq. (\ref{3}) simultaneously we can write
\ben
S_{3}=\int L_{1} \Phi d^4 x+\int L_{2} \sqrt{-g} d^4 x.
\label{5}
\een
This generalized formulation is known as the Weak version of NGVE theory. In Eq. (\ref{5}) the shift symmetry is applied only to $L_1$. $L_{1}$ and $L_{2}$ are independent of $\varphi_{a}$. To maintain infinite dimensional symmetry ($\varphi_{a}\rightarrow \varphi_{a}+f_{a}(L_1)$, where $f_{a}(L_1)$ is an arbitrary function of $L_1$), the mixing of $\Phi$ and $\sqrt{-g}$ like $\Phi^2/\sqrt{-g}$ are avoided \cite{Guendelman3}. Now, following \cite{Guendelman3}, the dynamics of an interacting scalar field ($\phi$) with the action (\ref{5}) is given below 
\ben
L_{1}=-\frac{1}{\k}R+\half g^{\mu\nu} \p_{\mu} \phi\p_{\nu}\phi-V(\phi)
\label{6}
\een
and 
\ben
L_{2}=U(\phi),
\label{7}
\een
where $\k=8\pi G=1$, $V(\phi)$ and $U(\phi)$ are two arbitrary potentials. 
In the above equations (\ref{5})--(\ref{7}) the measure scalar field ($\varphi_{a}$) and the interacting scalar field ($\phi$) are considered as independent variables. For arbitrary choice of $U(\phi)$ and $V(\phi)$ we can write by varying the measure fields ($\varphi_{a}$)
\ben
A^{\mu}_{a}\p_{\mu}L_{1}=0
\label{8}
\een
where $A^{\mu}_{a}=\e^{\mu\nu\a\b}\e_{abcd}\p_{\nu}\varphi_{b}\p_{\a}\varphi_{c}\p_{\b}\varphi_{d}$. Since $A^{\mu}_{a}\p_{\mu}\varphi_{a'}=(\delta a a'/4)\Phi$, it  is evident that $\det (A^{\mu}_{a})=(4^{-4}/4!)\Phi^{3}\neq 0$ if $\Phi \neq 0$. Hence if $\Phi \neq 0$ then $\p_{\mu}L_{1}=0$ i.e.,
\ben
L_{1}=-\frac{1}{\k}R+\half g^{\mu\nu}\p_{\mu}\phi\p_{\nu}\phi-V=M~(constant).
\label{9}
\een
The global scale invariance gives us the choice of $V(\phi)$ and $U(\phi)$ as \cite{Guendelman3}
\ben
V(\phi)=f_{1}e^{\a\phi},~~~~U(\phi)=f_{2}e^{2\a\phi},
\label{10}
\een
where $\a,~f_{1}$ and $f_{2}$ are constants.
To obtain the equation of motion (EoM) we can vary the action in Eq. (\ref{5}) concerning the metric $g_{\mu\nu}$ and obtain
\ben
\Phi\Big(\frac{-1}{\k}R_{\mu\nu}+\half \p_{\mu}\phi\p_{\nu}\phi \Big)-\half \sqrt{-g}U(\phi)g_{\mu\nu}=0.
\label{11}
\een
Reminding that $R=g^{\mu\nu}R_{\mu\nu}$ and considering Eq. (\ref{9}) we can get from Eq. (\ref{11})
\ben
M+V(\phi)-\frac{2U(\phi)}{\chi}=0.
\label{12}
\een
where $\chi=\Phi/\sqrt{-g}$, which is invariant under continuous general coordinate transformations, instead of the scalar density $\Phi$ \cite{Guendelman0}. This constraint equation (\ref{12}) gives us the expression of $\chi$ in terms of the potentials only 
\ben
\chi=\frac{2U(\phi)}{M+V(\phi)}.
\label{13}
\een

Using Eq. (\ref{10}) we can express $\chi$ as a function of scalar field ($\phi$) only
\ben
\chi=\frac{2f_{2}e^{2\alpha \phi}}{M+f_{1}e^{\alpha \phi}}.
\label{14}
\een

\subsection{Formation of a new geometry}

Following \cite{Guendelman3}, to construct the new geometry we modify the metric as the metric of Einstein conformal frame \cite{Hawking2}

\ben
\bar{g}_{\mu\nu}=\chi g_{\mu\nu},
\label{15}
\een
where $\chi$ is the conformal factor, and it is given in Eq. (\ref{13}) or (\ref{14}). The inverse metric of this geometry becomes
\ben
\bar{g}^{\mu\nu}=\frac{1}{\chi}g^{\mu\nu}.
\label{16}
\een

Note that,  it is evident from Eqs. (\ref{15}) and (\ref{16}), $\chi$ cannot be zero for the existence of the inverse metric (\ref{16}). If $\chi=1$ or $\chi=constant$, then the modified geometry ($\bar{g}_{\mu\nu}$) converts into the usual geometry $(g_{\mu\nu})$. Also, $(M+V(\phi))$ cannot be zero as $\chi$ can not be infinity (Eq. \ref{13}). So, we discarded these values of $\chi$ for our study.

Our study focuses on the new geometry ($\bar{g}_{\mu\nu}$) and its relationship to it, as specified in equations (\ref{15}) and (\ref{16}). Guendelman et al's NGVE theory \cite{Guendelman1, Guendelman3} clearly explains how the geometry is created. We aim to construct the RE using relations (\ref{15}) or (\ref{16}) and investigate cosmic possibilities based on the new formation of the modified RE ({\it viz.} below Eq. \ref{30}). Remember that the RE is a geometrical equation that is independent of any other gravitational theory and is not frame-dependent \cite{Kar, Dadhich}. Also, note that the RE is a completely scalar equation. So there is no confusion as to which frame (original or Einstein) we utilized.

Now we have to construct the Christoffel symbol based on the new metric ($\bar{g}_{\mu\nu}$). Using the definition of the Christoffel symbol, we can write
\ben
\bar{\Gamma}_{\nu\l}^{\mu}&&=\frac{1}{2}\g^{\mu\sg}\Big[\p_{\nu} \g_{\l\sg}+\p_{\l}\g_{\nu\sg}-\p_{\sg}\g_{\nu\l}\Big].
\label{17}
\een
The barred quantities are defined in modified geometry ($\bar{g}_{\mu\nu}$). From now on, all the quantities in modified geometry will be expressed in barred notations. 


The covariant derivative can be expressed as
\ben
\bar{\nb}_{\nu}v^{\mu}=\p_{\nu}v^{\mu}+\bar{\Gamma}_{\nu\l}^{\mu}v^{\l}.
\label{19}
\een


We can see that this newly formed geometry is dependent on $\chi$ (\ref{14}) and its derivatives. Due to the global invariance property of $\Phi$ it has been chosen to be a total derivative by Guendelman et al. \cite{Guendelman3}. This property allows $\Phi$ to overcome the global effect or surface contribution but it leaves the contribution of interaction inside the geometry (\ref{17}--\ref{20}). In our case, the dependency of the modified geometry has been transferred to the potentials $U(\phi)$ and $V(\phi)$. Hence this theory becomes a potential dependent theory.

The effective Einstein field equation (EEFE) \cite{Guendelman3} in this geometry gives
\ben
\R_{\mu\nu}-\half \bar{g}_{\mu\nu}\R=\frac{\k}{2}\T_{\mu\nu}^{eff}(\phi)
\label{20}
\een
where
\ben
\T^{eff}_{\mu\nu}(\phi)=\p_{\mu}\phi\p_{\nu}\phi-\half \bar{g}_{\mu\nu}\p_{\a}\phi\p_{\b}\phi\bar{g}^{\a\b}+\bar{g}_{\mu\nu}V_{eff}(\phi)
\label{21}
\een
and using Eqs. (\ref{9}), (\ref{10}) $\&$ (\ref{13}) 
\ben
V_{eff}(\phi)=\frac{1}{4U(\phi)}(V+M)^2=\frac{U(\phi)}{\chi^2} =\frac{1}{4f_{2}}(f_{1}+Me^{-\a\phi})^2.
\label{22}
\een
As we have the choice to transform $\phi\to-\phi$, we can always choose $\a>0$ \cite{Guendelman3}.\\

At this juncture, we wish to emphasize a particular aspect: the point particle model of matter in four dimensions ($D=4$) within the framework of Two Measure Theory (TMT) can be articulated such that the modified measure of matter, which interacts with the matter Lagrangian, fulfills the following condition:
\begin{equation}
g^{\mu \nu }\frac{\partial(L_{m})}{\partial g^{\mu \nu }}-(L_{m})=0.
\label{con}
\end{equation}
This enables the examination of point particle dynamics in conjunction with geodesic motion. The constraint that facilitates the resolution of the measure and the scalar field equation, thereby producing geodesic motion for point particles within the framework of TMT theory, remains unchanged in this instance, as the matter does not possess a direct correlation with the scalar field. This is attributable to the fact that a diverse array of actions is permissible for the freely falling point particle, which can be evaluated within the context of general relativity. In the context of four-dimensional space-time characterized by the metric $g_{\mu\nu}$, the standard action is expressed as $S=-m\int F(y)ds$, where $y=g_{\mu\nu}\frac{dX^{\mu}}{ds}\frac{dX^{\nu}}{ds}$ and $s$ is determined to be an affine parameter except if $F=\sqrt{y}$, which is the case of re-parametrization invariance. In TMT model, we must take $S_{m}=-m\int L_{part}\Phi d^{4}x$ with $L_{part}=-m\int ds\frac{\delta^{4}(x-X(s))}{\sqrt{-g}}F(y(X(s)))$ where $\int L_{part}\sqrt{-g}d^{4}x$ would be the action of a point particle in 4 dimensions in the usual theory. For the selection $F=y$, the constraint denoted as (\ref{con}) is satisfied.  In contrast to the scenario observed in general relativity, differing selections of $F$ result in non-equivalent theories. In general terms, the theory aligns with general relativity in the gauge $\chi=1$ provided that the matter Lagrangian is independent of curvature and inherently satisfies the constraint; this condition is met when the local Einstein symmetry is maintained. For a comprehensive discourse, please refer to \cite{Guendelman4}.\\

Before we go any further, we'd want to briefly explain the distinctions between studying the geodesic equation, the geodesic deviation equation and the Raychaudhuri equation.

The geodesic equation explores the trajectory of an individual particle or light beam inside spacetime, articulated by the equation $\frac{d^2 x^{\mu}}{d\tau^2}+\Gamma^{\mu}_{\nu\rho}\frac{dx^{\nu}}{d\tau}\frac{dx^{\rho}}{d\tau}=0$. This pertains to the trajectory $x^{\mu}(\tau)$ of an individual geodesic and integrates spacetime curvature indirectly via the Christoffel symbols $\Gamma^{\mu}_{\nu\rho}$. The geodesic equation may be used to the study of free particle motion, gravitational lensing, etc.

The geodesic deviation equation delineates the relative motion of adjacent geodesics, including tidal effects via the separation vector $\eta^{\mu}$. The equation is expressed as $\frac{D^2\eta^{\mu}}{d\tau^2}=-R^{\mu}_{\nu\rho\sigma}u^{\nu}\eta^{\rho}u^{\sigma}$, wherein the Riemann tensor $R^{\mu}_{\nu\rho\sigma}$ is explicitly present. This equation examines the relative behavior of two geodesics and is used to investigate tidal forces, gravitational waves and geodesic stability, among other phenomena.

The Raychaudhuri equation analyzes the development of a family (congruence) of geodesics or non-geodesics. It's written as $\frac{d\theta}{d\tau}=-\frac{1}{3}\theta^2-\sigma_{\mu\nu}\sigma^{\mu\nu}+\o_{\mu\nu}\o^{\mu\nu}-R_{\mu\nu}u^{\mu}u^{\nu}$, including expansion $\theta$, shear $\sigma_{\mu\nu}$, and vorticity $\o_{\mu\nu}$. Spacetime curvature is represented by the Ricci tensor $R_{\mu\nu}$. The equation investigates the collective behavior of a congruence and has applications in gravitational focusing, singularity theorems, and cosmology, among others.\\

{\it We clearly state that our present study does not include the investigation of geodesic motion or the geodesic deviation equation, as it concentrates on the Raychaudhuri equation within the framework of NGVE theory, which elucidates the evolution of congruence of geodesics rather than the path of an individual geodesic; furthermore, it does not address the geodesic or geodesic deviation equations, which are regarded as distinct areas of study.}

\section{Modified Raychaudhuri Equation in the context of NGVE theory}
The commutation relation between the covariant derivatives in this modified geometry ($\bar{g}_{\mu\nu}$) gives the Riemann curvature of this geometry along the vector field $v^{\a}$ \cite{Poisson, Blau, Chandrasekhar}
\ben
(\bar{\nb}_{\mu}\bar{\nb}_{\nu}-\bar{\nb}_{\nu}\bar{\nb}_{\mu})v^{\a}=\bar{R}_{~\rho\mu\nu}^{\a}v^{\rho},
\label{23}
\een
where $\bar{R}_{~\rho\mu\nu}^{\a}$ is the Riemann tensor of the new geometry.

Contracting $\a$ with $\mu$ and then with $v^{\nu}$ we get
\ben
v^{\nu}\bar{\nb}_{\mu}\bar{\nb}_{\nu}v^{\mu}-v^{\nu}\bar{\nb}_{\nu}\bar{\nb}_{\mu}v^{\mu}=\R_{\rho\nu}v^{\rho}v^{\nu}
\label{24}
\een
where $\R_{\rho\nu}$ is the Ricci tensor of modified geometry.
Multiplying both sides by $(-1)$ and rearranging by parts, we get
\ben
v^{\nu}\bar{\nb}_{\nu}\bar{\nb}_{\mu}v^{\mu}+\bar{\nb}_{\mu}v^{\nu}\bar{\nb}_{\nu}v^{\mu}-\bar{\nb}_{\mu}(v^{\nu}\bar{\nb}_{\nu}v^{\mu})=-\R_{\rho\nu}v^{\rho}v^{\nu}.
\label{25}
\een
This is basically the RE in the new geometry. If $v^{\mu}$ is affinely parametrized vector field, then $v^{\nu}\bar{\nb}_{\nu}v^{\mu}=0$ (geodesic equation) and the third term vanishes. $\bar{\nb}_{\mu}v^{\mu}$ is the scalar expansion ($\Th$) of new geometry. It should be noted that selecting an affinely connected vector field in the derivation of the Raychaudhuri equation ensures consistency with the geometric structure of spacetime and simplifies the analysis of congruences of geodesics. The consideration of affinely parametrized vector field ensures that the vector field in modified geometry and usual geometry would be equivalent i.e., $\bar{v}_{\mu}\equiv v_{\mu}$. 

From Eq. (\ref{25}) we can write
\ben
v^{\nu}\bar{\nb}_{\nu}\Th+\bar{\nb}_{\mu}v^{\nu}\bar{\nb}_{\nu}v^{\mu}=-\R_{\rho\nu}v^{\rho}v^{\nu}
\label{26}
\een
Therefore, the {\it modified RE of this geometry} becomes
\ben
\frac{d\bar{\Theta}}{d\bar{s}}+(\bar{\nb}_{\mu}v^{\nu})(\bar{\nb}_{\nu}v^{\mu})=-\R_{\rho\nu}v^{\rho}v^{\nu}
\label{27}
\een
where $\bar{s}$ is an affine parameter in modified geometry (\ref{15}). 
The second term of Eq. (\ref{27}) on the left-hand side may be divided into three components: modified scalar expansion ($\bar{\Theta}$), modified symmetric shear ($\bar{\sigma}$), and modified antisymmetric rotation ($\bar{\omega}$) as
\ben
(\bar{\nb}_{\mu}v^{\nu})(\bar{\nb}_{\nu}v^{\mu})=2\bar{\sigma}^2-2\bar{\omega}^2+\frac{1}{3}\bar{\Theta}^2
\label{27A}
\een
where $\bar{\sigma}_{\mu\nu}=\frac{1}{2}(\bar{\nb}_{\beta}v^{\a}+\bar{\nb}_{\a}v_{\b})-\frac{1}{3}\bar{\Theta}\bar{h}_{\a\b},~\bar{\omega}_{\a\b}=\frac{1}{2}(\bar{\nb}_{\beta}v^{\a}-\bar{\nb}_{\a}v_{\b}),~\bar{\nb}_\b v_\a=\bar{\sigma}_{\a\b}+\bar{\o}_{\a\b}+\frac{1}{3}\bar{\Theta} \bar{h}_{\a\b},~\bar{h}_{\a\b}=\bar{g}_{\a\b}-v_{\a}v_{\b}
$.
Therefore, the modified RE in terms of the above parameters of this geometry can be written as
\ben
\frac{d\bar{\Theta}}{d\bar{s}}+(2\bar{\sg}^2-2\bar{\omega}^2+\frac{1}{3}\bar{\Theta}^2)=-\R_{\rho\nu}v^{\rho}v^{\nu}
\label{27B}
\een

Being a geometrical identity, the {\it form of RE} (\ref{27}) or (\ref{27B}) is the same in any geometry. Therefore, we cannot understand the alterations induced by the modification of geometry ($\bar{g}_{\mu\nu}$) from the perspective of the modified RE (\ref{27}) or (\ref{27B}). Understanding the interactions is possible just via external observation. To study the significance of this modified RE (\ref{27}) we extract the interaction terms, which is dependent on the conformal factor $\chi$.

It is important to acknowledge that theoretically, ``wherever there are vector fields describing a physical or geometrical quantity, there must be corresponding Raychaudhuri equations" \cite{Kar}, and, as noted in \cite{Hensh}, ``an important fact about those equations is that they are purely geometrical and do not assume any theory of gravity, providing the freedom to use any theory to fix the geometry and study the evolution of congruence."

The second term of Eq. (\ref{27}) has the transformation relation as
\ben
&&(\bar{\nb}_{\mu}v^{\nu})(\bar{\nb}_{\nu}v^{\mu})\nn\\
&&=(\nb_{\mu}v^{\nu})(\nb_{\nu}v^{\mu})+\frac{1}{2\chi}\Big[2\theta v^{\l}(\p_{\l}\chi)\Big]-\frac{1}{2\chi}\Big[\nb_{\mu}v^{\nu}(\p^{\mu}\chi)v_{\nu}+\nb^{\sg}v^{\mu}(\p_{\sg}\chi)v_{\mu}\Big]+\frac{1}{4\chi^2}\Big[3(\p_{\mu}\chi)(\p_{\nu}\chi)v^{\mu}v^{\nu}\nn\\
&&-2(\p_{\mu}\chi)(\p^{\mu}\chi)v_{\l}v^{\l}-2(\p_{\l}\chi)v_{\nu}(\p^{\nu}\chi)v^{\l}+4(\p_{\l}\chi)v^{\l}(\p_{\l}\chi)v^{\l}+(\p^{\mu}\chi)v_{\mu}(\p^{\nu}\chi)v_{\nu}\Big]
\label{28}
\een
where the first part of the R.H.S. corresponds to the usual case \cite{Ray1, Ray2, Ray3} as 
\ben
(\nb_{\mu}v^{\nu})(\nb_{\nu}v^{\mu})=2\sg^2-2\omega^2+\frac{1}{3}\theta^2.
\label{29}
\een
Consider an affinely parametrized vector field for the usual gravitational case, and we set \cite{Blau, Poisson, Ray1, Ray2, Ray3} $v^\a\nb_\a v^\b=0$, $\t=\nb_\a v^\a$ is the usual scalar expansion, symmetric shear $\sigma_{\a\b}=\frac{1}{2}\Big(\nb_\b v_\a+\nb_\a v_\b\Big)-\frac{1}{3}\t h_{\a\b}$,  
antisymmetric rotation $\o_{\a\b}=\frac{1}{2}\Big(\nb_\b v_\a-\nb_\a v_\b\Big)$, $2\sigma^{2}=\sigma_{\a\b}\sigma^{\a\b}$, $2\o^{2}=\o_{\a\b}\o^{\a\b}$ and 
 $\nb_\b v_\a=\sigma_{\a\b}+\o_{\a\b}+\frac{1}{3}\t h_{\a\b}$ with the three-dimensional hypersurface metric $h_{\a\b}=g_{\a\b}-v_\a v_\b$.

Simplifying the transformation relation in Eq. (\ref{28}) we get the RE in this modified geometry from Eq. (\ref{27})
\ben
\frac{d\bar{\Theta}}{d\bar{s}}+\Big(2\sg^2-2\omega^2+\frac{1}{3}\theta^2\Big)+\frac{1}{\chi^2}\Big[(\p_{\l}\chi)v^{\l}(\p_{\l}\chi)v^{\l}\Big]=-\R_{\rho\nu}v^{\rho}v^{\nu}
\label{30}
\een
In the modified RE (\ref{30}), the second term within the first bracket plays the conventional function of the original RE, while the other terms are alterations that arise from the emergence of new geometry under the NGVE theory. It is important to mention that splitting the modified RE (\ref{30}) into conventional gravitational components and additional terms is essential for understanding the RE of NGVE theory. It identifies deviations from standard RE and gives a clear physical interpretation of new contributions (such as vacuum energy or fields). This distinction helps in the identification and analysis of unique dynamical effects while maintaining a connection to standard GR for theoretical and empirical comparisons.

\section{Cosmological Analysis through the Modified Raychudhuri Equation}

Considering the background gravitational metric to be flat Friedmann-Lemaitre-Robertson-Walker (FLRW) \cite{Blau, Weinberg2}, the line element can be written as
\ben
ds^{2}=dt^{2}-a^{2}(t)\sum_{i=1}^{3} (dx^{i})^{2},
\label{31}
\een
where $a(t)$ is the scale factor. So, the modified line element under the NGVE theory can be written using \ref{15} as
\ben
d\bar{s}^2=\chi dt^2-\chi a^2 \sum_{i=1}^{3} (dx^{i})^{2}.
\label{31A}
\een

Now we want to calculate the scalar expansion ($\bar{\Theta}$) of modified theory. From Eq. (\ref{15}) we can calculate
\ben
\det(\bar{g}_{\mu\nu})=\det(\chi g_{\mu\nu})=\chi^4 \det(g_{\mu\nu}).
\label{32}
\een
Therefore,
\ben
\sqrt{-\bar{g}}=\chi^2\sqrt{-g}=\chi^2 a^3,
\label{33}
\een
where $\bar{g}$ is the determinant of $\bar{g}_{\mu\nu}$.
We consider the vector field $(v^{\mu})$ is in comoving Lorentz frame i.e., $v^{\mu}=(1,0,0,0)$. So from the definition of the scalar expansion \cite{Poisson}, $\Th$  can be written as 
\ben
\Th=\bar{\nb}_{\a}v^{\a}=\frac{1}{\sqrt{-\bar{g}}}\p_{\a}(\sqrt{-\bar{g}}~v^{\a})=3\frac{\dot{a}}{a}+\frac{2\dot{\chi}}{\chi}= \theta+\frac{2\dot{\chi}}{\chi}
\label{34}
\een
where $\theta=3\frac{\dot{
a}}{a}$ is the usual scalar expansion of the flat FLRW spacetime. 

We now consider the interacting matter field ($\phi$) that relies only on time i.e., $\phi\equiv \phi(t)$, indicating that the conformal factor ($\chi$) is likewise time-dependent, as can be readily shown from Eq. (\ref{14}). The homogeneity of the scalar field $\phi$ can be assured by the use of the flat-homogeneous FLRW metric (\ref{31}). In other words, to maintain the cosmological principle, we consider our system to be homogenous and isotropic.

Considering the vector field $v^{\mu}=(1,0,0,0)$, the $(00)$ component of the modified RE (\ref{30}) i.e., the timelike RE of this geometry, can be written as
\ben
\frac{d\bar{\Theta}}{d\bar{s}}+(2\sg^2-2\omega^2+\frac{1}{3}\theta^2)+\Big(\frac{\dot{\chi}}{\chi}\Big)^2=-\R_{00}
\label{35}
\een

Using Eqs. (\ref{15}), (\ref{17}) and (\ref{31A}), we can summarize the non-zero connection coefficients of the modified geometry (\ref{15}) in FLRW background (\ref{31}) as
\ben
\bar{\Gamma}_{00}^{0}=\frac{\dot{\chi}}{\chi},~~~\bar{\Gamma}_{11}^{0}=a\dot{a}+\frac{a^{2}\dot{\chi}}{2\chi},~~~\bar{\Gamma}_{22}^{0}=a\dot{a}+\frac{a^{2}\dot{\chi}}{2\chi},~~~\bar{\Gamma}_{33}^{0}=a\dot{a}+\frac{a^{2}\dot{\chi}}{2\chi}\nn\\
\bar{\Gamma}_{01}^{1}=\frac{\dot{a}}{a}+\frac{\dot{\chi}}{2\chi}=\bar{\Gamma}_{10}^{1},~~~\bar{\Gamma}_{02}^{2}=\frac{\dot{a}}{a}+\frac{\dot{\chi}}{2\chi}=\bar{\Gamma}_{20}^{},~~~\bar{\Gamma}_{03}^{3}=\frac{\dot{a}}{a}+\frac{\dot{\chi}}{2\chi}=\bar{\Gamma}_{30}^{3}.
\label{36}
\een

Now to investigate the cosmological implications of the modified RE (\ref{30}) in this geometry ($\bar{g}_{\mu\nu}$), we have to evaluate $\R_{00}$. The Ricci  tensor ($\R_{\mu\nu}$) in this geometry can be written as,
\ben
\R_{\mu\nu}=\partial_{\rho}\bar{\Gamma}_{\mu\nu}^{\rho}-\partial_{\mu}\bar{\Gamma}_{\nu\rho}^{\rho}+\bar{\Gamma}_{\rho\sigma}^{\rho}\bar{\Gamma}_{\mu\nu}^{\sigma}-\bar{\Gamma}_{\mu\sigma}^{\rho}\bar{\Gamma}_{\rho\nu}^{\sigma}
\label{36A}
\een
The transformation relation between $\R_{\mu\nu}$ of modified metric and $R_{\mu\nu}$ of usual metric can be written as
\ben
\R_{\mu\nu}=R_{\mu\nu}-\frac{1}{\chi}\nb_{\mu}\nb_{\nu}\chi+\frac{1}{2\chi^2}\nb_{\mu}\chi\nb_{\nu}\chi+\frac{1}{2\chi}g_{\mu\nu}\square \chi-\frac{1}{4\chi^2}g_{\mu\nu}g^{\rho\sigma}\nb_{\rho}\chi\nb_{\sigma}\chi
\label{36B}
\een
Using the non-zero connection coefficients (\ref{36}) in Eq. (\ref{36A}), the $(00)$ component of the Ricci tensor can be written as, 
\ben
\R_{00}=-3\frac{\ddot{a}}{a}-\frac{3}{2}\frac{\dot{a}}{a}\frac{\dot{\chi}}{\chi}-\frac{3}{2}\frac{\ddot{\chi}}{\chi}+\frac{3}{2}\Big(\frac{\dot{\chi}}{\chi}\Big)^2
\label{37}
\een

As the standard method for studying cosmic behaviors using the RE, we use the ideal fluid model and can define the effective energy-momentum tensor as:
\ben
\T^{eff}_{\a\b}(\phi)=(\bar{\rho}+\bar{p})v_{\a}v_{\b}-\g_{\a\b}\bar{p}.
\label{38}
\een
The `bar' over $\rho$ and $p$ has been used to indicate that the energy density ($\bar{\rho}$) and pressure ($\bar{p}$) in the geometry defined by $\bar{g}_{\mu\nu}$ (\ref{15}) are different than the usual energy density and pressure of the usual geometry of $g_{\mu\nu}$.

Therefore, using the EEFE (\ref{20}) with the energy-momentum tensor (\ref{38}) for the perfect fluid,  we have the relation
\ben
\R_{\a\b}v^{\a}v^{\b}= 4\pi G (\bar{\rho}+3\bar{p}).
\label{39}
\een
From the left-hand side of Eq. (\ref{20}) and using (\ref{38}), we can calculate
\ben
8\pi G\bar{\rho}=3\frac{\dot{a}^2}{a^2}+\frac{3}{4}\frac{\dot{\chi}^2}{\chi^2}+3\frac{\dot{a}}{a}\frac{\dot{\chi}}{\chi}
\label{40}
\een
and 
\ben
8\pi G\bar{p}=-2\frac{\ddot{a}}{a}+\frac{3}{4}\frac{\dot{\chi}^2}{\chi^2}-2\frac{\dot{a}}{a}\frac{\dot{\chi}}{\chi}-\frac{\ddot{\chi}}{\chi}-\frac{\dot{a}^2}{a^2}.
\label{41}
\een
Note that the Eqs. (\ref{40}) and (\ref{41}) are nothing but the well-known first and second (modified) Friedmann equations for this geometry. 
Combining the above two Eqs. (\ref{40}) and (\ref{41}), we can write
\ben
4\pi G(\bar{\rho}+3\bar{p})=-3\frac{\ddot{a}}{a}+\frac{3}{2}\frac{\dot{\chi}^2}{\chi^2}-\frac{3}{2}\frac{\dot{a}}{a}\frac{\dot{\chi}}{\chi}-\frac{\ddot{\chi}}{\chi}=\bar{R}_{00}.
\label{42}
\een
Using Eq.(\ref{35}), the modified RE become
\ben
\frac{d\bar{\Theta}}{d\bar{s}}=-2\sg^2+2\o^2-\frac{1}{3}\theta^2-\Big(\frac{\dot{\chi}}{\chi}\Big)^2-4\pi G(\bar{\rho}+3\bar{p}).
\label{43}
\een
It should be noted that if we consider the conformal coupling factor $\chi$ to be unity, then the Eq. (\ref{43}) reduces to its original RE \cite{Ray2, Ray3, Kar, Poisson} as
\ben
\frac{d\bar{\Theta}}{d\bar{s}}=-2\sg^2+2\o^2-\frac{1}{3}\theta^2-4\pi G(\rho+3p)\equiv \frac{d\theta}{ds}.
\label{44}
\een

On the other hand, from the definition of the scalar expansion (\ref{34}), we have the rate of change of the scalar expansion with respect to the cosmological time parameter in the comoving frame, ($\frac{dt}{d\bar{s}}=1$) \cite{Das} is: 
\ben
\frac{d\bar{\Theta}}{d\bar{s}}=\frac{d\bar{\Theta}}{dt}\frac{dt}{d\bar{s}}=3\frac{\ddot{a}}{a}-3\Big(\frac{\dot{a}}{a}\Big)^2+2\frac{\ddot{\chi}}{\chi}-2\Big(\frac{\dot{\chi}}{\chi}\Big)^2.
\label{45}
\een

Now comparing the Eqs. (\ref{43}) and (\ref{45}), we obtained the acceleration equation (another form of the modified RE) of this geometry as follows: 
\ben
&& \frac{\ddot{a}}{a}=-\frac{2}{3}\sg^{2}+\frac{2}{3}\o^{2}-\frac{2}{3}\frac{\ddot{\chi}}{\chi}+\frac{1}{3}\Big(\frac{\dot{\chi}}{\chi}\Big)^2-\frac{4\pi G}{3}(\bar{\rho}+3\bar{p})\nn\\
\Rightarrow && \frac{\ddot{a}}{a}=\Big[-\frac{2}{3}\sg^{2}+\frac{2}{3}\o^{2}-\frac{4\pi G}{3}(\bar{\rho}+3\bar{p})\Big]+f(\chi)
\label{46}
\een
where, 
\ben
f(\chi)=\frac{1}{3}\Big(\frac{\dot
{\chi}}{\chi}\Big)^2-\frac{2}{3}\frac{\ddot{\chi}}{\chi}.
\label{47}
\een
From (\ref{46}), we can say that the acceleration of the universe can be expressed in terms of shear ($\sigma$), rotation ($\o$), matter contribution ($\bar{\rho}$ and $\bar{p}$) and the interaction terms ($f(\chi)$).\\ 

To further analyze, we need to clearly describe the $f(\chi)$ term using Eq. (\ref{14}), where we treat the interacting field ($\phi$) as a function of time. Therefore, we can evaluate Eq. (\ref{47}) with the help of Eq. (\ref{14}) as
\ben
f(\chi)\equiv f(\phi,\dot{\phi},\ddot{\phi})=-\frac{\alpha  \Big[2 \ddot{\phi} \Big(f_1^2 e^{2 \alpha  \phi}+3 f_1 M e^{\alpha  \phi}+2 M^2\Big)+\alpha  \dot{\phi}^2 \Big(f_1^2 e^{2 \alpha  \phi}+2 f_1 M e^{\alpha  \phi}+4 M^2\Big)\Big]}{3 \Big(f_1 e^{\alpha  \phi}+M\Big)^2}.
\label{48}
\een

Observe that Eqs. (\ref{43}) or (\ref{46}) represents the RE within the new conformal geometry (\ref{15}) in the FLRW background and this equation may be used to investigate the cosmic implications of this geometry. We studied two instances based on whether $f(\chi)$ is zero or non-zero. Furthermore, if the vorticity term ($\o^2$) is zero in order to maintain the ideal fluid model, we noted that the third bracket on the right-hand side of Eq. (\ref{46}) is negative owing to the positivity of ($\bar{\rho}+3\bar{p}$) \cite{Ray3, Kar, Hawking1}. \\

It should also be noted that before subsection C, we will discuss the existence of two types of effective energy-momentum tensors (\ref{21} and (\ref{38}), which provide us the constraint equations for the equality of these two tensors.

\subsection{Analysis for $f(\chi)\neq 0$}
To view the impact of the scenarios of the cosmology derived from the new geometry in Eq. (\ref{46}), we may have $f(\chi) \neq 0$ (setting $\o^2=0$). Upon the condition of positivity, negativity or equality three cases can arise from Eq. (\ref{46}).\\

{\bf Case A:} If 
\ben
f(\chi)>\frac{2}{3}\sg^{2}+\frac{4\pi G}{3}(\bar{\rho}+3\bar{p})
\label{49}
\een
then we have $\frac{\ddot{a}}{a}>0$ from (\ref{46}). This indicates the notion of an {\it expanding} universe without singularity. The above condition (\ref{49}) tells us that if the contribution of $\chi$ dominates over the shear and matter part, we get a situation where the geodesics are moving away from each other with acceleration. The usual Big Bang theory tells us a different story of geodesic evolution, whereas our condition shows the opposite nature. Similar behavior has been achieved in \cite{Choudhury, Das} in different scenarios. Note that, to satisfy this condition, $f(\chi)$ should be a positive quantity.\\

{\bf Case B:} If
\ben
f(\chi)<\frac{2}{3}\sg^{2}+\frac{4\pi G}{3}(\bar{\rho}+3\bar{p})
\label{50}
\een
then Eq. (\ref{46}) gives $\frac{\ddot{a}}{a}<0$ which indicates the deceleration. All terms on the right-hand side of (\ref{46}) (in the absence of $\o^2$) contribute to the universe's {\it collapsing} scenarios, consistent with the typical use of the Raychaudhuri equation in the FLRW metric \cite{Ray3, Kar, Hawking1}. In this case, $f(\chi)$ could be both positive or negative along with the condition (\ref{50}).\\

{\bf Case C:} If we set
\ben
f(\chi)=\frac{2}{3}\sg^{2}+\frac{4\pi G}{3}(\bar{\rho}+3\bar{p}).
\label{51}
\een
This condition makes the R.H.S. of Eq. (\ref{46}) zero i.e., $\frac{\ddot{a}}{a}=0$ which indicates a {\it steady-state} model of the universe \cite{Hoyle, Bondi,Hoyle1, Hoyle2, Hoyle3, Hoyle4}. The steady-state model has been discarded in contemporary cosmology due to observational evidence. In this case, $f(\chi)$ should be positive.

The solutions obtained this far are derived from the RE of the new geometry (\ref{15}) in the flat homogeneous FLRW background. In the subsequent section of this work, we will examine a specific case to analyze the model.

\subsubsection{Behavior of $f(\chi)$ through the Solution of the Equation of Motion}

It is evident from the three situations (Case A, Case B, and Case C) mentioned above that $f(\chi)$ must be either positive or negative in order to maintain the requirements stated in Eqs. (\ref{49}), (\ref{50}), and (\ref{51}). Therefore, it is crucial to identify whether $f(\chi)$ is positive or negative. To understand the nature of $f(\chi)$, we need to explicitly define the scalar field $\phi$ with time, since $\chi$ depends on $\phi$ (\ref{14}). The equation of motion of the scalar field in modified geometry ($\bar{g}_{\mu\nu}$) can be written as \cite{Guendelman3}:
\ben
\frac{1}{\sqrt{-\bar{g}}}\partial_{\mu}\Big(\bar{g}^{\mu\nu}\sqrt{-\bar{g}}\partial_{\nu}\phi\Big)+V'_{eff}(\phi)=0
\label{52}
\een
where prime denotes the derivative with respect to the argument within the parenthesis.

Using Eqs. (\ref{15}) and (\ref{16})  we can write
\ben
\frac{1}{\chi^2\sqrt{-g}}\partial_{\mu}\Big(\frac{1}{\chi}g^{\mu\nu}\chi^2\sqrt{-g}\partial_{\nu}\phi\Big)+V'_{eff}(\phi)=0.
\label{53}
\een
Using Eqs. (\ref{14}) and (\ref{22}), we have
\ben
V'_{eff}=\frac{-\alpha M}{2f_{2}}e^{-\alpha\phi}(f_{1}+Me^{-\alpha\phi}).
\label{54}
\een
Using the expressions of Eqs. (\ref{14}) and (\ref{33}) in Eq. (\ref{53}), we get
\ben
(\frac{M+f_{1}e^{\a\phi}}{2f_{2}e^{2\a\phi}})^2\frac{1}{a^3}\partial_{t}(\frac{2f_{2}e^{2\a\phi}a^3\dot{\phi}}{M+f_{1}e^{\a\phi}})+V'_{eff}(\phi)=0.
\label{55}
\een
Now solving Eq. (\ref{55}) we get
\ben
\dot{\phi}(t)=\frac{(M+f_{1}e^{\a\phi})e^{C_{1}-2\a\phi}}{a^3}
\label{56}
\een
where $C_{1}$ is an integration constant. This delineates the relationship between the scalar field ($\phi$) and the scale factor ($a(t)$). This applies equally to all types of scale factors. Also, we can express the scale factor ($a$) in terms of the scalar field ($\phi$) as
\ben
a(t)=\Big[\frac{(M+f_{1}e^{\a\phi})e^{C_{1}-2\a\phi})}{\dot{\phi}}\Big]^{1/3}
\label{57}
\een

To get an explicit expression for $\phi$ in terms of $t$, it is necessary to examine the various forms of the scale factor $a(t)$ as follows:

\begin{center}
 $\blacksquare$ {\bf Study of Special Cases}
\end{center}

\begin{center}
{\it i. Exponential Scale Factor}
\end{center}

To solve Eq. (\ref{56}), we may consider a particular scale factor, which is an exponential scale factor 
\ben
a(t)=e^{H_{0}t}
\label{58}
\een
where $H_{0} (=\frac{\dot{a}}{a})$ is the Hubble parameter for a particular epoch. Note that, the acceleration of the cosmos during inflation in the early universe \cite{Mukhanov,Peebles,Liddle} 
and later times when dark energy is the main impact \cite{Weinberg2} can both be described by this specific scale factor (\ref{58}). 

Performing integrations and solving Eq. (\ref{56}) we get
\ben
-\frac{1}{Be^{-\a\phi}}+\frac{A\a\phi}{B^2}+\frac{A\ln (B+Ae^{-\a\phi})}{B^2}=\a\frac{e^{-3H_{0}t}}{3H_{0}}+C_2
\label{59}
\een
where $A=Me^{C_1}$, $B=f_{1}e^{C_1}$ with $C_1$ and $C_{2}$ are integration constants. From Eq. (\ref{59}), $\phi(t)$ can be simplified as
\ben
\phi(t)=\frac{1}{\a}\ln \Big[\frac{1}{B}\Big(A \Big(-W\Big(-\frac{e^{\frac{B^2 K}{A}+\frac{\alpha  B^2 e^{-3 H t}}{3 A H}-1}}{A}\Big)\Big)-A\Big)\Big]
\label{60}
\een
where $W(x)$ is the Lambert $W$ function also called the omega function or product logarithm \cite{Gray,Corless, Majumder}, $K$ is an integration constant. By using Eq. (\ref{60}), we may evaluate the characteristics of $f(\chi)$. It is important to take care of certain constants and model parameters in this instance. Table (\ref{Table1}) describes all model parameters and constants along with their respective sources.
\begin{table}[h]
\begin{center}
\resizebox{12cm}{!}{
\begin{tabular}{ |c|c|c| } 
\hline
\text{Parameter} &  \text{Source} & \text{Types} \\
\hline
$K$ & Expression of $\phi$, Eq. (\ref{60}) &  integration constants \\
$A~\&~B$ & Relation between scalar field and time, Eq. (\ref{59}) & integration constants \\
$\a$  & Potentials $U(\phi)$ and $V(\phi)$, Eq. (\ref{10}) & model parameters \\
  $f_{1}~\&~f_{2}$  & Potentials $U(\phi)$ and $V(\phi)$, Eq. (\ref{10}) & model parameters \\
    $M$  & Lagrangian $L_{1}$, Eq. (\ref{9}) & model parameter \\
     $H_{0}$  & Scale factor, Eq. (\ref{58}) & model parameter \\
  \hline
\end{tabular}}
\end{center}
\caption{Constants and model parameters in this theory}
\label{Table1}
\end{table}

It has been verified that a negative $K$ cannot yield any plot and that $H_{0}$, the Hubble parameter for a certain epoch, cannot be negative. The scalar field ($\phi$) implies that $\a$ is selected to be positive based on its transformation characteristic. Furthermore, from Eq. (\ref{48}), it is evident that $f_{2}$ is not present in the equation. Consequently, we left with $A, B, f_{1}$, and $M$.

\begin{figure*}[h]
\begin{minipage}[b]{0.4\linewidth}
\centering
 \begin{subfigure}[b]{0.9\textwidth}
    \includegraphics[width=6cm]{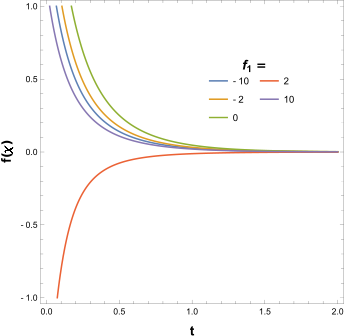}
       \caption{Variation of $f(\chi)$ with $t$ for varying $f_{1}$}
        \label{Fig1a}
    \end{subfigure}
\end{minipage}
\hspace{2cm}
\begin{minipage}[b]{0.4\linewidth}
\centering
 \begin{subfigure}[b]{0.9\textwidth}
    \includegraphics[width=6cm]{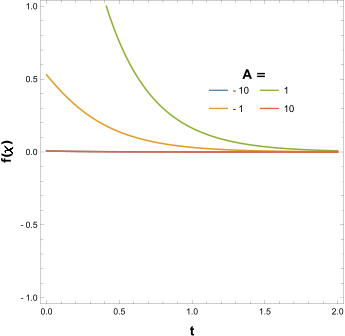}
        \caption{ 
Variation of $f(\chi)$ with $t$ for varying A}
        \label{Fig1b}
    \end{subfigure}
\end{minipage}
\caption{Variation of $f(\chi)$ with $t$ for (a) different $f_{1}$ and (b) different $A$}
\label{Fig1}
\end{figure*}

\begin{figure*}[h]
\begin{minipage}[b]{0.4\linewidth}
\centering
 \begin{subfigure}[b]{0.9\textwidth}
    \includegraphics[width=6cm]{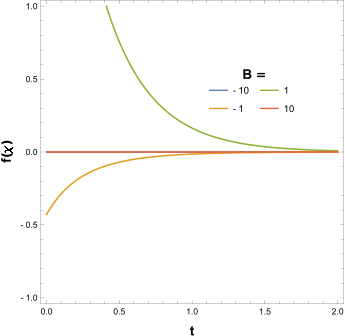}
        \caption{ 
Variation of $f(\chi)$ with $t$ for varying B}
        \label{Fig2a}
    \end{subfigure}
\end{minipage}
\hspace{2cm}
\begin{minipage}[b]{0.4\linewidth}
\centering
 \begin{subfigure}[b]{0.9\textwidth}
    \includegraphics[width=6cm]{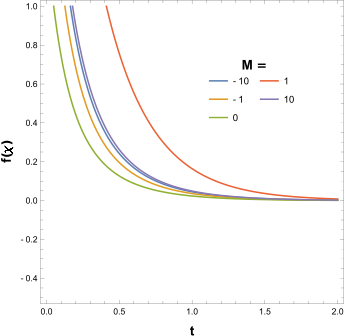}
        \caption{ 
Variation of $f(\chi)$ with $t$ for varying M}
        \label{Fig2b}
    \end{subfigure}
\end{minipage}
\caption{Variation of $f(\chi)$ with $t$ for different (a) $B$ and (b) $M$}
\label{Fig2}
\end{figure*}

Figure (\ref{Fig1a}) represents the variation of the function $f(\chi)$ with time for different values of $f_{1}~(-10,-2,0,2,10)$. A negative nature is seen around $f_{1}=2$, otherwise for all the values of $f_{1}$, $f(\chi)$ shows positive nature. 

Figure (\ref{Fig1b}) shows the variation of $f(\chi)$ with time where $A$ has been varied. $f(\chi)$ can positive and zero depending upon the values of $A$. No negative nature has been observed here. It has been noted, that $A=0$ does not provide any plot for $f(\chi)$. The Eqs. (\ref{59}) and (\ref{60}) supports this fact too.

Figure (\ref{Fig2a}) shows the variation of $f(\chi)$ with time where $B$ is varied. Here also $B=0$ does not provide any plot for $f(\chi)$ which can be verified by Eq. (\ref{60}). Only for $B=-1$ negative nature of $f(\chi)$ has been achieved. The plots for $B=10$ and $B=-10$ are overlapped.

Figure (\ref{Fig2b}) is the variation of $f(\chi)$ with time for different values of $M$ ($-10,-1,0,1,10$). For all possible values of $M$ the nature of $f(\chi)$ is positive.

In Figs. (\ref{Fig1b}) and (\ref{Fig2a}), it is seen that in some instances, $f(\chi)=0$ for all values of $t$, while in all cases, $f(\chi)$ attains a value of zero at a certain time. Consequently, it is essential to examine the situation of $f(\chi)=0$, which is addressed later in this article.

The usual RE demonstrates a collapsing universe in the absence of the vorticity terms. But, in our study, we see that the presence of the interactive terms involving the scalar field in the modified RE produces three different cases based on the conditions imposed on the scalar field. A single theoretical framework can give rise to models of the universe that exhibit expansion (\ref{49}), collapse (\ref{50}), or a steady-state configuration (\ref{51}). Comparable results were previously demonstrated by Das et al. \cite{Das} within the context of K-essence geometry, a non-conformal theoretical approach. In this study, we examine the modified RE within the framework of conformal geometry, grounded in the principles of the NGVE theory.

\begin{center}
    {\it ii. Power Law Scale Factor}
\end{center}

Next, we choose the scale factor as a power law type as
\ben
a=a_{0}t^{m},~~m>0,
\label{61}
\een
where $a_{0}$ is a constant, which is considered to be $1$. Note that, here $t=0$ is not the present time and $a_{0}$ is not the present value of the scale factor. Here, $a_{0}$ is just a scaling factor for the scale parameter. This type of scale factor has been used in the context of cosmology and astrophysics \cite{Kaplinghat, Weinberg2}.  Recently, the power law cosmology has been used to explain most of the distinguished attributes of evolution in cosmology including the thermodynamic behavior of modified gravity theories \cite{Singh}.

This consideration gives us the solution of $\phi(t)$ as
\ben
\phi(t)=\frac{1}{\a}\ln\Big[\frac{1}{B} \Big\{-A+A \Big(-W\Big(-\frac{\exp \Big(-1-\frac{\alpha  B^2 \Big(\frac{t^{1-3 m}}{a_0^3 (1-3 m)}+K_1\Big)}{A}\Big)}{\alpha  A B}\Big)\Big)\Big\}\Big]
\label{62}
\een

\begin{figure*}[h]
\centering
    \includegraphics[width=7cm]{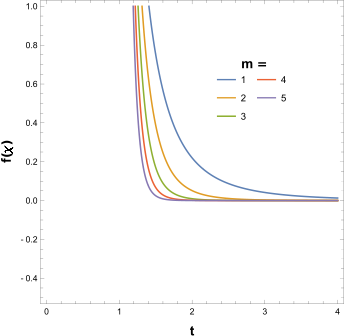}
       \caption{Variation of $f(\chi)$ with $t$ for varying $m$ for Power Law scale factor}
\label{Fig3}
\end{figure*}

Using this solution (\ref{62}) in Eq. (\ref{48}) to examine the characteristics of $f(\chi)$, we obtained findings similar to those seen with the exponential scale factor (\ref{58}). The constants and model parameters $A,~B,~\a,~f_{1},~f_{2}$ and $M$ are present in both the solution of Eqs. (\ref{60}) and (\ref{62}). Only difference is, $H_{0}$ in the Eq. (\ref{58}) is replaced by $m$ in Eq. (\ref{61}). We here only show the variation of $f(\chi)$ with $m$ (Fig. \ref{Fig3}). It has been seen that for all values of $m$, we get a positive nature of $f(\chi)$. This positive nature has an important contribution in the case of the conditional study of Eqs. (\ref{49}), (\ref{50}) and (\ref{51}).

\subsection{Analysis for $f(\chi)=0$}
From Eq. (\ref{46}), it is evident that when $f(\chi)=0$, a specific form of modified geometry is obtained, wherein the conformal factor $\chi$ does not exhibit direct dependence; however, it is indirectly manifested through the modified density ($\bar{\rho}$) and pressure ($\bar{p}$), as demonstrated in Eqs. (\ref{40}) and (\ref{41}). Furthermore, we get an explicit time-dependent solution for $\chi$ from the constraint $f(\chi)=0$. So, applying the condition $f(\chi)=0$ in Eq. (\ref{47}), we get
\ben
\chi=\frac{c_1{}^2 t^2}{4 c_2}+c_1 t+c_2
\label{63}
\een
where $c_{1},~c_{2}$ are arbitrary integration constants. A particular scenario occurs when $c_{1}=0~(c_{2}\neq 0)$, resulting in $\chi=c_{2}~(constant)$, which indicates the equivalence (\ref{15}) between NGVE geometry ($\bar{g}_{\mu\nu}$) and conventional geometry ($g_{\mu\nu}$). Consequently, there is no new information under this condition ($c_1=0$), so we exclude this solution ($\chi=constant$) from our analysis. On the other hand, if $c_{1}$ and $c_{2}$ both are non-zero, we encounter a scenario where $f(\chi)$ equals zero while $\chi\neq 0$, indicating the presence of interaction terms via $\bar{\rho}$ and $\bar{p}$.  Eventually, these expressions become (using Eqs. (\ref{40}) and (\ref{41}))
\ben
\bar{\rho}=\frac{3}{(c_1 t+2 c_2)^2}\Big[\frac{2 c_1 (c_1 t+2 c_2) \dot{a}}{a}+\frac{a^2 c_1^2+(c_1 t+2 c_2)^2 \dot{a}^2}{a^2}\Big]
\label{64}
\een
and 
\ben
\bar{p}=\frac{a c_1^2-2 (c_1 t+2 c_2)^2 \ddot{a}}{a (c_1 t+2 c_2)^2}-\frac{\dot{a}^2}{a^2}-\frac{4 c_1 \dot{a}}{c_1 t a+2 c_2 a}.
\label{65}
\een
with $8\pi G=1$. Under the assumption that $f(\chi)=0$, these findings (Eqs. (\ref{64}) and (\ref{65})) might lead to a specific type of modified gravity.

Now we define the equation of state (EoS) parameter ($\Omega =\frac{\bar{p}}{\bar{\rho}}$) of this NGVE geometry for this special case. We may express the EoS parameter using the energy density ($\bar{\rho}$) and pressure ($\bar{p}$) of this specific situation, as stated in Eqs. (\ref{64}) and (\ref{65}), respectively, as
\ben
\Omega=\frac{-(c_1 t+2 c_2){}^2 \dot{a}^2-2 a (c_1 t+2 c_2) \Big((c_1 t+2 c_2) \ddot{a}+2 c_1 \dot{a}\Big)+c_1{}^2 a^2}{3 \Big((c_1 t+2 c_2) \dot{a}+c_1 a\Big){}^2}.
\label{66}
\een
We use separate $\Omega$ values for different epochs to solve the scale factor for this unique instance with $c_{3}$ and $c_{4}$ being other integration constants.
\begin{table}[H]
\begin{center}
\resizebox{15cm}{!}{
\begin{tabular}{ |c|c|c| } 
\hline
\text{Value of $\Omega$} &  \text{Epoch} & \text{Solutions of scale factor (a(t))} \\
\hline
$-2$ & Phantom era &  $c_4 \exp \Big[-\frac{1}{3} c_1 \Big(\frac{3 \ln (c_1 t+2 c_2)}{c_1}+\frac{2 \ln ((c_1 t+2 c_2){}^2+c_3)}{c_1}\Big)\Big]$ \\
$-1$ & Dark Energy Era & $\frac{c_4 e^{\frac{1}{2} c_1 c_3 t^2+2 c_2 c_3 t}}{c_1 t+2 c_2}$ \\
$0$  & Dust era & $c_4 \exp \Big[\frac{1}{3} c_1 \Big(\frac{2 \ln \Big(1+c_3 (c_1 t+2 c_2){}^2\Big)}{c_1}-\frac{3 \ln (c_1 t+2 c_2)}{c_1}\Big)\Big]$ \\
  $0.5$  & Early universe & $c_4 \exp \Big[-\frac{1}{9} c_1 \Big(\frac{9 \ln (c_1 t+2 c_2)}{c_1}-\frac{4 \ln \Big(1+c_3 (c_1 t+2 c_2){}^2\Big)}{c_1}\Big)\Big]$ \\
    $1$  & Stiff Fluid era & $c_4 \exp \Big[\frac{1}{3} c_1 \Big(\frac{\ln \Big(1+c_3 (c_1 t+2 c_2){}^2\Big)}{c_1}-\frac{3 \ln (c_1 t+2 c_2)}{c_1}\Big)\Big]$ \\
  \hline
\end{tabular}}
\end{center}
\caption{Scale factor for different epochs of the universe}
\label{Table2}
\end{table}

Table (\ref{Table2}) shows the solution of the scale factor derived from Eq. (\ref{66}) for the condition $f(\chi)=0$ where $\chi$ has the expression of (\ref{63}) using different values of $\O$. The parameter 
$\O$ correspond to different cosmological eras: $\O=-2$ represents the phantom era, $\O=-1$ corresponds the dark energy-dominated era, $\O=0$ signifies the dust-dominated era, $\O=0.5$ represents the early universe, and $\O=1$ characterizes the stiff fluid era. The corresponding acceleration in terms of $\frac{\ddot{a}}{a}$ has been plotted below. The factor $\frac{\ddot{a}}{a}$ is associated with the second Friedmann equation in FLRW spacetime. In standard geometry, it relates to the trace of the energy-momentum tensor. For the case of $f(\chi)=0$, the interaction terms manifest through the components of the energy-momentum tensor. Therefore, studying the acceleration parameter using $\frac{\ddot{a}}{a}$ provides valuable insights into the dynamics of the system.

\begin{figure*}[h]
\begin{minipage}[b]{0.4\linewidth}
\centering
 \begin{subfigure}[b]{0.9\textwidth}
    \includegraphics[width=6cm]{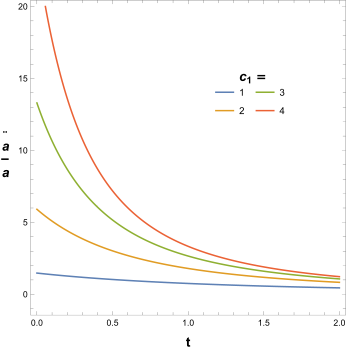}
       \caption{Variation of $\frac{\ddot{a}}{a}$ for $\O=-2$ (phantom era)}
        \label{Fig4a}
    \end{subfigure}
\end{minipage}
\hspace{2cm}
\begin{minipage}[b]{0.4\linewidth}
\centering
 \begin{subfigure}[b]{0.9\textwidth}
    \includegraphics[width=6cm]{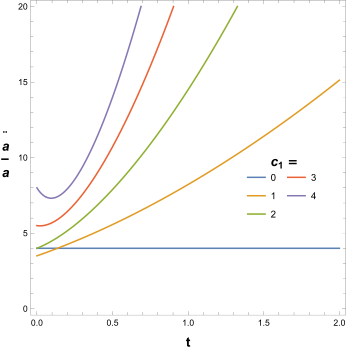}
        \caption{ 
Variation of $\frac{\ddot{a}}{a}$ for $\O=-1$ (dark energy era)}
        \label{Fig4b}
    \end{subfigure}
\end{minipage}
\caption{Variation of $\frac{\ddot{a}}{a}$ with $t$ for (a) $\O=-2$ and (b) $\O=-1$ [$f(\chi)=0)$ case]}
\label{Fig4}
\end{figure*}

\begin{figure*}[h]
\begin{minipage}[b]{0.4\linewidth}
\centering
 \begin{subfigure}[b]{0.9\textwidth}
    \includegraphics[width=6cm]{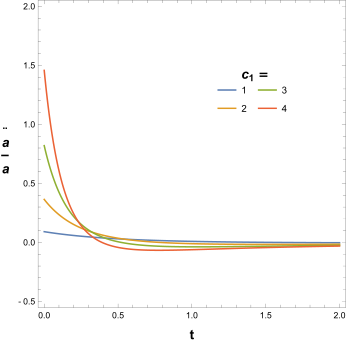}
       \caption{Variation of $\frac{\ddot{a}}{a}$ for $\O=0$ (dust era)}
        \label{Fig5a}
    \end{subfigure}
\end{minipage}
\hspace{2cm}
\begin{minipage}[b]{0.4\linewidth}
\centering
 \begin{subfigure}[b]{0.9\textwidth}
    \includegraphics[width=6cm]{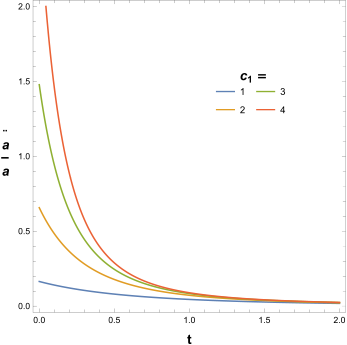}
        \caption{ 
Variation of $\frac{\ddot{a}}{a}$ for $\O=0.5$ (early universe)}
        \label{Fig5b}
    \end{subfigure}
\end{minipage}
\caption{Variation of $\frac{\ddot{a}}{a}$ with $t$ for (a) $\O=0$ and (b) $\O=0.5$ [$f(\chi)=0)$ case]}
\label{Fig5}
\end{figure*}

\begin{figure*}[h]
    \includegraphics[width=6cm]{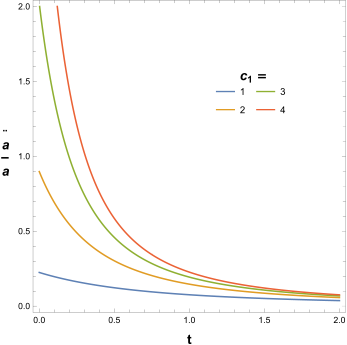}
       \caption{Variation of $\frac{\ddot{a}}{a}$ for $\O=1$ (stiff fluid era) [$f(\chi)=0$ case]}
\label{Fig6}
\end{figure*}

Figs. (\ref{Fig4a}) and (\ref{Fig4b}) represent the variation of $\frac{\ddot{a}}{a}$ with cosmic time when $c_{1}$ is varied for $\O=-2$ (phantom era) and $\O=-1$ (dark energy era) respectively. Figs. (\ref{Fig5a}) and  (\ref{Fig5b}) respectively illustrate  the variation of $\frac{\ddot{a}}{a}$ with time for $\O=0$ (dust era) and $\O=0.5$ (early universe). Fig. (\ref{Fig6}) presents  the variation of $\frac{\ddot{a}}{a}$ with time for $\O=1$ (stiff fluid era).

Studying these graphs we get that in the phantom era the acceleration drops exponentially to the zero value as time progresses (Fig. \ref{Fig4a}) whereas, in the dark energy-dominated era the acceleration increases rapidly with time (Fig. \ref{Fig4b}). This indicates the accelerated expansion of the universe as suggested by the observations \cite{Riess,Perlmutter,Komatsu,Planck,Planck1,Planck2}. In the dust era, a decreasing nature of the acceleration factor ($\frac{\ddot{a}}{a}$) has been seen, like the phantom case (Fig. \ref{Fig5a}) but the slopes of the graphs are greater than the phantom case. Moreover, for large value of the integration constant ($c_{1}$) it has been seen that the value of $\frac{\ddot{a}}{a}$ goes below zero and then increases after a period of time and becomes constant at a zero value. A similar decrease has been seen in the case of the early universe (Fig. \ref{Fig5b}) and the stiff fluid era (Fig. \ref{Fig6}) too. For all these graphs we see a nature of exponential increase or decrease of the term ($\frac{\ddot{a}}{a}$). Observing the values of the scale factor given in Table (\ref{Table2}) it is clear that $\frac{\ddot{a}}{a}$ has an exponential behavior. Several articles support the fact that the universe may have an exponential expansion throughout its evolution \cite{Frieman,Nemiroff,Smeulders}. A zero acceleration phase can be obtained in Brans-Dicke theory \cite{Kim}. The authors of \cite{Kim} showed that the BD scalar field-dominated era is a period of zero acceleration that serves as a transitional period between the decelerating matter-dominated era and the accelerating phase of the present universe. In our case Fig. (\ref{Fig5a}) shows that the decreasing nature of $\frac{\ddot{a}}{a}$ is ended at zero acceleration in a different context. In matter or dust-dominated epoch a slower rate of acceleration has been observed by the decreasing nature of $\frac{\ddot{a}}{a}$. In the case of the early universe (\ref{Fig5b}) the acceleration starts from a high value which may be the indication of the inflationary expansion. Then we get a decreasing nature of the acceleration, which tell us that the acceleration becomes slower after a certain period of time. In each case, except the dark energy period, the acceleration terms reach at zero value which may be the indication of the transition periods of the corresponding era.
\\

It is important to note that we have two ways of calculating the components of the energy-momentum tensor from Eq. (\ref{20}) under the perfect fluid assumptions (\ref{38}). We may get $\bar{\rho}$ and $\bar{p}$ by equating the left-hand side and right-hand side of Eq. (\ref{20}) with Eq. (\ref{38}), separately. From equating the left-hand side of Eq. (\ref{20}) with Eq. (\ref{38}), we found the expression of $\bar{\rho}$ and $\bar{p}$ in Eqs. (\ref{40}) and (\ref{41}). On the other hand, if we calculate the components of the energy-momentum tensor from Eq. (\ref{21}) when the matter field behaves as perfect-fluid-like (\ref{38}), we get
\ben
\bar{\rho}=\half \K+\frac{U(\phi)}{\chi}=\half \K+\bar{V}(\phi)
\label{67}
\een
and 
\ben
\bar{p}=\half \K-\frac{U(\phi)}{\chi}=\half \K-\bar{V}(\phi)
\label{68}
\een
where $\bar{V}(\phi)=\frac{U(\phi)}{\chi}$. These expressions of $\bar{\rho}$ and $\bar{p}$ are similar to what we obtained in the case of any standard scalar field cosmology ({\it viz.} quintessence) \cite{Weinberg2}. Thus, we can conclude that adopting the perfect fluid model within the NGVE theory avoids any inconsistencies. In a single theory, we cannot use two distinct energy-momentum tensors in relation to $\bar{\rho}$ in both (\ref{40}) and (\ref{67}), and the equation for $\bar{p}$ in both (\ref{41}) and (\ref{68}) must be identical. Therefore we get two constraint equations as
\ben
\half \K+\frac{U(\phi)}{\chi}=\frac{1}{8\pi G}\Big[3\frac{\dot{a}^2}{a^2}+\frac{3}{4}\frac{\dot{\chi}^2}{\chi^2}+3\frac{\dot{a}}{a}\frac{\dot{\chi}}{\chi}\Big]
\label{69}
\een
and 
\ben
\half \K-\frac{U(\phi)}{\chi}=\frac{1}{8\pi G}\Big[-2\frac{\ddot{a}}{a}+\frac{3}{4}\frac{\dot{\chi}^2}{\chi^2}-2\frac{\dot{a}}{a}\frac{\dot{\chi}}{\chi}-\frac{\ddot{\chi}}{\chi}-\frac{\dot{a}^2}{a^2}\Big]
\label{70}
\een

With the aid of Eq. (\ref{14}), a relationship between the scalar field and the scale factor can be found using these constraint equations (\ref{69}) and (\ref{70}). However, due to the complexity of the equation, no analytical solution to $\phi$ can be deduced. On the contrary, we use the equation of motion of the NGVE theory to solve the scalar field. To demonstrate that the ideal fluid model may be applied to NGVE theory, we represent the equations (\ref{69}) and (\ref{70}) here. 

It should also be noted that Eq. (\ref{67}) to Eq. (\ref{70}) may give us the tools to study the cosmological behavior of the universe, which is associated with the EEFE and the perfect fluid model consideration, which would be an independent work in the NGVE theory that does not concern about the Raychaudhuri equation. However, in this article, we are not interested in the study of the cosmological analysis of the model from the effective Einstein's equation. Here, we constrain ourselves to investigating the modified RE and the cosmological aspects of the modified RE only.

\subsection{Focusing theorem}

According to the Focusing theorem \cite{Poisson}, congruences that are initially converging ($\theta<0$) will converge faster in the future, while those that are initially diverging ($\theta>0$) will diverge slower. In accordance with Einstein's General Relativity, the Raychaudhuri equation predicts, if a congruence of timelike geodesics is hypersurface orthogonal ($\o_{\a\b}=0$), then the fulfillment of the strong energy condition (SEC) ($R_{\a\b}v^{\a}v^{\b}\geq 0$) indicates $\frac{d\theta}{ds}\leq 0$. In other words, the growth rate of the congruences of geodesics will be hampered as it progress through the spacetime. Validation of SEC says that gravity is attractive. Looking at Eq. (\ref{43}) we can say that this scenario of geodesic convergence is applicable in our case if $\R_{\a\b}v^{\a}v^{\b}\geq 0$ i.e., $4\pi G (\bar{\rho}+\bar{p})\geq 0$. In Eq. (\ref{43}) all the terms in the R.H.S are negative ($\o^2=0$) including the interaction term and $\frac{d\bar{\Theta}}{d\bar{s}}\leq 0$. As a result we can say that focusing theorem holds true in our geometry.

But besides this focusing condition, we also achieved a condition of non-focusing when we consider the variation of modified expansion (\ref{45}). The sign of the terms $\frac{\ddot{a}}{a}$ and $\frac{\ddot{\chi}}{\chi}$ are undetermined. Therefore we cannot be sure about $\frac{d\bar{\Theta}}{d\bar{s}}\leq 0$. Furthermore, it is mentioned that after a specific congruence of geodesics in the gravitational field, the expansion $\bar{\Theta}$ provides a naturally coordinate-independent way to determine whether the free-falling matter is converging or diverging, where $\bar{\Theta}>0$ indicates divergence and $\bar{\Theta}<0$ indicates convergence \cite{Albareti2012, Das}. The authors of \cite{Albareti2012} showed that the space-time contribution to the Raychaudhuri equation in an accelerated expanding universe is positive for the basic congruence, suggesting a non-focusing of the congruence of geodesics, according to a simple calculation from the metric of a FLRW cosmological model.

\subsubsection{Caustic formation}
Let's represent Eq. (\ref{45}) as
\ben
\frac{d\bar{\Theta}}{d\bar{s}}=\frac{d\theta}{dt}+2\frac{\ddot{\chi}}{\chi}-2\Big(\frac{\dot{\chi}}{\chi}\Big)^2
\label{71}
\een
where $\frac{d\theta}{dt}=3\frac{\ddot{a}}{a}-3\Big(\frac{\dot{a}}{a}\Big)^2$, which is the usual scalar expansion of the flat FLRW spacetime. Therefore, for $\o^2=0$, using Eqs. (\ref{71}), (\ref{45}) and (\ref{43}), we have
\ben
\frac{d\theta}{dt}+\frac{1}{3}\theta^2=-2\sg^2-4\pi G(\bar{\rho}+3\bar{p})+F(\chi)
\label{72}
\een
where $F(\chi)=-2\frac{\ddot{\chi}}{\chi}+(\frac{\dot{\chi}}{\chi})^2$. A similar type of investigation can be found in \cite{Das} for a different context. The authors have studied the modified Raychaudhuri equation in a non-canonical theory named K-essence. In Eq. (\ref{72}) we accumulate the interacting terms of the RE in the factor $F(\chi)$ which determines the caustic solutions. In \cite{Das} this factor was dependent on the non-canonical scalar field whereas in our case this dependency is on the conformal factor $(\chi)$. Thus, the investigation of caustic solutions yields the same type of consequences as in Ref. \cite{Das}, but in a completely different geometry.

The nature of $F(\chi) (\neq 0)$ is undetermined, hence following \cite{Das}, we take two cases 

{\it Case I:} If
\ben
2\sg^2+4\pi G(\bar{\rho}+3\bar{p})\geq F(\chi)
\label{73}
\een

From Eq. (\ref{72}) it can be said that the rate of change of scalar expansion is negative which means the scalar expansion can approach $-\infty$ for both positive and negative nature of $F(\chi)$. Similar to \cite{Das} this condition gives rise to a singular focus point. Consequently, the congruence will provide a caustic point where different geodesics converge. The occurrence of the singularity is called the spacetime future incompleteness or geodesic incompleteness \cite{Ray1,Ray2,Kar}.


{\it Case II:} If
\ben
F(\chi)> 2\sg^2+4\pi G(\bar{\rho}+3\bar{p}),
\label{74}
\een
in this scenario, the rate of expansion exceeds a negative value, resulting in a range that extends from a negative value to a positive infinity. As noted in \cite{Das}, this condition leads to two distinct cases. In the first case, $\frac{d\theta}{dt}$ lies between the negative value and zero, while in the second case, $\frac{d\theta}{dt}$ lies between zero and positive infinity. The first case corresponds to an initially converging congruence of geodesics, whereas the second case represents an initially diverging congruence. Depending upon the initial convergence of the geodesic a singular universe can be obtained whereas the initial divergence gives rise to a singularity-free universe.

The focusing of the congruence is not evident because there are conditions on whether $\frac{d\bar{\Theta}}{d\bar{s}}$ is positive or not for our situation. Firstly we showed, that there may be a propensity for the congruence to diverge as an accelerated expansion rather than gravitational collapse if the focusing theorem is not true under the condition stated in Eq. (\ref{49}). In another context, Das \cite{Das1} has also examined the prevention of the creation of the focussing as a semiclassical variant of the singularity theorem. Consequently, we can create several kinds of singularity-free universes \cite{Senovilla,Senovilla1,Maddox}. Secondly, we have shown that for initially diverging congruence, we obtain $\theta\rightarrow \infty$ which also corresponds to a non-singular universe \cite{Bhattacharyya}. On the other hand, possibility of converging geodesics i.e., the case of future incompleteness also has been seen in the same model [{\it viz.} Eqs. (\ref{50}) and (\ref{73})]. These are merely cosmological implications. A supernova explosion might occur in the case of non-focusing conditions. A white dwarf, neutron star, black hole, or in the case of gravitational collapse an interesting quark star could develop \cite{Choudhury2, Hensh}. These are only a few of the possible interpretations if we study astrophysics. The potential existence of either an accelerated expansion or a decelerated phase may be found on the basis of Eqs. (\ref{46}) and (\ref{72}), respectively \cite{Albareti2012}.

\section{Conclusion}

The NGVE theory, a unique physical framework, has the potential to unveil various gravitational phenomenon and contains an Einstein sector of solutions \cite{Guendelman0, Guendelman1}. There are various ways to expand research on NGVE theory. In this article, we investigate the cosmological implications of the universe through the well-known Raychaudhuri equation in the framework of NGVE theory.

The RE is a powerful tool in addressing fundamental questions in cosmology and astrophysics, including the singularity problem. Starting with a basic version of NGVE theory, we investigate the action (\ref{3}), which has two components. The first component involves the measure field ($\Phi$) coupled to the Lagrangian $L_{1}$ (\ref{6}), while the second component adopts the standard metric formulation with a scalar field potential (\ref{7}).

Following the approach in \cite{Guendelman3}, we employ the global scale invariance of the metric to define the potentials as exponential functions (\ref{10}).  Variations in the action provide the equations of motion (EoM) for conventional gravity (\ref{11}) and a constraint equation (\ref{12}). The function $\chi$ includes the combined contributions of the measure field and metric properties and is defined in terms of the scalar field potentials $U(\phi)$ and $V(\phi)$ (\ref{13}). Because these potentials rely entirely on the interacting scalar field $\phi$, the component $\chi$ may be written as a function of $\phi$. Consequently, the Lagrangian $L_{1}$ reduces to a constant $M$ (\ref{9}).

This formulation proposed by Guendelman \cite{Guendelman3} has been utilized to formulate a new geometry in terms of a modified metric, $\bar{g}_{\mu\nu}$ (\ref{15}). This modified metric is conformally related to the usual metric by the conformal factor ($\chi$) which has been deduced from the NGVE theory in Eq. (\ref{14}). Now we are standing in a geometry which is not defined by the usual metric rather by the conformal metric $\bar{g}_{\mu\nu}$. The connection coefficients (\ref{17}), covariant derivative (\ref{19}) and the Einstein field equation (\ref{20}) change accordingly with the help of the conformal metric $\bar{g}_{\mu\nu}$. The effective energy-momentum tensor can be expressed as Eq. (\ref{21}).

Based on the NGVE theory, we define the commutation relation between the covariant derivatives (\ref{23}) in the new geometry. The final form of the modified RE has been expressed in Eq. (\ref{30}). 

After formulating the structure of the modified RE, we delve into the study of cosmology regarding the modified RE of the NGVE geometry. In this process, we consider the background metric to be flat FLRW (\ref{31}). Under this consideration, the modified scalar expansion ($\bar{\Theta}$) (\ref{34}) has been deduced in terms of the scale factor ($a$) and the interaction terms, and the timelike modified RE takes the form of Eq. (\ref{35}). To evaluate the dynamics we calculate the $\R_{00}$ of the geometry in flat FLRW background. The matter contribution has been considered to be the perfect fluid type (\ref{38}) in the modified geometry. The energy density (\ref{40}) and pressure (\ref{41}) in this geometry have been calculated by solving the L.H.S. of the EEFE (\ref{20}). Therefore, we achieved the final form of timelike RE in Eq. (\ref{43}) to study the modified RE's cosmological implications. Another form of modified RE which is the acceleration equation has been evaluated in Eq. (\ref{46}) where the interaction terms are contained inside the function $f(\chi)$. It has been noted that the interaction terms present in the energy density ($\bar{\rho}$) and pressure ($\bar{p}$) have not been extracted. The expression of $f(\chi)$ (\ref{47}) is then studied under two conditions. One is $f(\chi)\neq 0$ and the other is $f(\chi)=0$.  

Studying the $f(\chi)\neq 0$ we get three conditions that give rise to the conditional existence of expanding (\ref{49}), collapsing (\ref{50}), and steady-state model (\ref{51}). The condition is determined by the interaction terms $f(\chi)$, hence a study of $f(\chi)$ is necessary. The expression of $f(\chi)$ is a function of $\phi$ therefore, to express it in terms of a more convenient physical parameter we solve the equation of motion of the NGVE theory and evaluate $\phi$ in terms of cosmic time ($t$). To do so, we consider two types of scale factors: one is the exponential scale factor (\ref{58}) other one is the power law scale factor (\ref{61}). For the exponential scale factor, we get the solution of $\phi$ in Eq. (\ref{60}) and for the power law scale factor, we get the solution $\phi$ in Eq. (\ref{62}). With these solutions, we have studied the nature of $f(\chi)$ graphically. From the Figs. (\ref{Fig1})--(\ref{Fig3}) we can see that $f(\chi)$ can be positive, negative or zero depending upon the model parameters mentioned in Table (\ref{Table1}). 

On the other hand, the study of the $f(\chi)=0$ case reveals a special type of geometry that $f(\chi)=0$ does not require $\chi=0$. As the interaction terms are present in modified energy density ($\bar{\rho}$) and pressure ($\bar{p}$), therefore the condition of $f(\chi)$ zero does not eliminate the interactions of the new geometry completely from the modified RE (\ref{46}). The condition $f(\chi)=0$ allows us to express $\chi$ as a function of time (\ref{63}) and hence we have computed the modified energy density and pressure for this special case in Eqs. (\ref{64}) and (\ref{65}). These give rise to the modified EoS parameter of this special case in Eq. (\ref{66}). Studying Eq. (\ref{66}) we see that this equation can be utilized to find the scale factor for different eras of the universe. Table (\ref{Table2}) represents all the solutions corresponding to the different epochs of the universe characterized by the $\O$ value of the relevant epochs. Thus the scale factors used to find the acceleration of the universe for different epochs and the acceleration in the form of $\ddot{a}/a$ have been plotted in Figs.(\ref{Fig4})--(\ref{Fig6}). In dark energy dominated era (Fig. \ref{Fig4b}) we see the acceleration increases exponentially with the evolution of time which supports the observations. In other cases like phantom era (Fig. \ref{Fig4a}), dust dominated era (Fig. \ref{Fig5a}), early universe era (Fig. \ref{Fig5b}) and stiff fluid era (Fig. \ref{Fig6}) the acceleration term ($\ddot{a}/a$) starts with a large value and approaches zero with the evolution of time which may be the indication of transition from the corresponding era. The decreasing nature of ($\ddot{a}/a$) also indicates that the acceleration is getting slower in these eras.

Lastly, we also studied the focusing theorem and the caustic formation in case of the timelike geodesics utilizing the modified RE. When the modified scalar expansion $\bar{\Theta}$ has been expressed in terms of the usual scalar expansion ($\theta$) (\ref{72}) we get two conditions: Case I gives rise to a caustic point where the geodesics are in far future can converge and produce a singularity. Case II contributes to two scenarios depending upon the initial nature of the geodesics. If the geodesics are convergent initially, singularity comes into play, whereas if the geodesics are divergent initially the expansion occurs in the future. This summarizes the focusing theorem of the modified RE.  

We would like to note at the end of the conclusion that this theory has a limitation, the complexity of the constraint equations stated in Eqs. (\ref{69}) and (\ref{70}). These equations are complicated to solve and beyond the scope of this study.  The Friedmann equations of the newly stated geometry [Eqs. \ref{40} and (\ref{41})] were used to examine the modified Raychaudhuri equation. We leave open the possibility that the constraint equations may provide new avenues for the study of cosmic behavior. However, in this work, we only focus on the investigation of some cosmological scenarios through the newly formed modified RE in the NGVE theory.

\vspace{0.3in}

{\bf Acknowledgement:}
G.M. would like to extend thanks to all the undergraduate, postgraduate, and doctoral students who significantly enriched him.\\

{\bf Conflicts of interest:} The authors declare no conflicts of interest.\\

{\bf Data availability:} There is no associated data with this article, and as such, no new data was generated or analyzed in support of this research.\\

{\bf Declaration of competing interest:}
The authors declare that they have no known competing financial interests or personal relationships that could have appeared to influence the work reported in this paper.\\

{\bf Declaration of generative AI in scientific writing:} The authors state that they do not support the use of AI tools to analyze and extract insights from data as part of the study process.\\


\begin{thebibliography}{99}



\bibitem{Ray1} 
A. Raychaudhuri, ``Relativistic Cosmology. I", Phys. Rev. {\bf 98}, 1123 (1955) \url{https://doi.org/10.1103/PhysRev.98.1123}


\bibitem{Heckmann1}
O. Heckmann and E. Schucking, ``Remarks on Newtonian Cosmology. I", Z. Astrophysik 38, 95 (1955) \url{https://adsabs.harvard.edu/pdf/1955ZA.....38...95H}

\bibitem{Ray2} 
A. Raychaudhuri, ``Relativistic and Newtonian cosmology", Z. Astrophys. {\bf 43}, 161 (1957) \url{https://adsabs.harvard.edu/full/1957ZA.....43..161R}

\bibitem{Ray3} 
A. K. Raychaudhuri, S. Banerji and A. Banerjee, ``General Relativity, Astrophysics and Cosmology", (Astronomy \& Astrophysics Library, Springer (2003) \url{https://link.springer.com/book/9780387406282}

\bibitem{Kar} 
S. Kar and S. Sengupta, ``The Raychaudhuri equations: A brief review", Pramana - J. Phys. {\bf 69}, 49 (2007) \url{https://api.semanticscholar.org/CorpusID:119438891}

\bibitem{Dadhich}
N. Dadhich, ``Derivation of the Raychaudhuri Equation", 	arXiv:gr-qc/0511123 (2005)\url{
https://doi.org/10.48550/arXiv.gr-qc/0511123}

\bibitem{Hensh}
S. Hensh, S. Liberati, ``Raychaudhuri equations and gravitational collapse in Einstein-Cartan theory'', Phys. Rev. D {\bf 104}, 084073 (2021) \url{https://doi.org/10.1103/PhysRevD.104.084073}

\bibitem{Chakraborty1}
M Chakraborty, S Chakraborty, ``Implications of Raychaudhuri equation and geodesic focusing in interacting two fluid systems'', Eur. Phys. J. C {\bf 85},114 (2025)  \url{https://doi.org/10.1140/epjc/s10052-025-13850-6}

\bibitem{Chakraborty2}
M Chakraborty, S Chakraborty, ``Classical and quantum analysis of gravitational singularity from Raychaudhuri equation'', Phys. Lett. A, {\bf 525}, 129883 (2024) \url{https://doi.org/10.1016/j.physleta.2024.129883}
 
\bibitem{Hawking1}
S. W. Hawking, ``Singularities in the universe", Phys. Rev. Lett. {\bf 17}, 444 (1966) \url{https://doi.org/10.1103/PhysRevLett.17.444}


\bibitem{Hawking2}
S.W. Hawking and G.F.R. Ellis, ``The Large Scale Structure of Space-time" (Cambridge University Press, 1999) \url{https://doi.org/10.1017/9781009253161}.



\bibitem{Penrose1}
R. Penrose, ``Gravitational collapse and space-time singularities", Phys. Rev. Lett. {\bf 14}, 57 (1965) \url{ https://doi.org/10.1103/PhysRevLett.14.57}

\bibitem{Hawking3}
S. W. Hawking, ``Occurrence of singularities in open universes", Phys. Rev. Lett. {\bf 15}, 689 (1965) \url{https://doi.org/10.1103/PhysRevLett.15.689}
\bibitem{Penrose2}
R. Penrose, R.D. Sorkin, E. Woolgar, ``A Positive Mass Theorem Based on the Focusing and Retardation of Null Geodesics", arXiv e-prints, (1993) \url{https://ui.adsabs.harvard.edu/link_gateway/1993gr.qc.....1015P/arxiv:gr-qc/9301015}

\bibitem{Choudhury}
 S. G. Choudhury, et al., ``Reconstruction of 
$f(R)$ gravity models for an accelerated universe using the Raychaudhuri equation",  Month. Not. Roy. Astron. Soc., {\bf 485}, 5693. (2019) \url{https://doi.org/10.1093/mnras/stz731}.

\bibitem{Das}
S. Das et al., ``Raychaudhuri Equation in K-essence Geometry: Conditional Singular and Non-Singular Cosmological Models", Fortschr. Phys, {\bf 71}, 2200193 (2023) \url{https://doi.org/10.1002/prop.202200193}


\bibitem{Poisson}
E. Poisson, Relativist’s Toolkit, Cambridge University Press, Cambridge (2004) \url{https://doi.org/10.1017/CBO9780511606601}

\bibitem{Panda}
A. Panda, D. Gangopadhyay, and G. Manna, `` Form Invariance of Raychaudhuri Equation in the Presence of Inflaton-Type Fields",  Fortschr. Phys. {\bf 72}, 2400134, (2024) \url{https://doi.org/10.1002/prop.202400134} 

\bibitem{Panda2}
A. Panda, D. Gangopadhyay, and G. Manna, `` NEC violation in $f(R,T)$ gravity in the context of a non-canonical theory via modified Raychaudhuri equation", Astropart. Phys., {\bf 165}, 103059, (2025), \url{https://doi.org/10.1016/j.astropartphys.2024.103059} 

\bibitem{Straumann}
N. Straumann. “The Mystery of the Cosmic Vacuum Energy Density and the Accelerated
 Expansion of the Universe”. Eur. J. Phys., {\bf 20}(6):419, (1999) \url{https://iopscience.
 iop.org/article/10.1088/0143-0807/20/6/307}.
 
\bibitem{Carroll}
S. M. Carroll. “The Cosmological Constant”. Living Rev. Relativ., {\bf 4}(1):1–56, (2001) \url{https://doi.org/10.12942%2Flrr-2001-1}

\bibitem{Martin}
J. Martin. “Everything You Always Wanted to Know About the Cosmological Constant
 Problem (but were Afraid to Ask)”. C. R. Phys., {\bf 13}(6-7):566–665, (2012) \url{https:
 //doi.org/10.1016/j.crhy.2012.04.008}.

\bibitem{Peri} 
L. Perivolaropoulos and F. Skara. “Challenges for $\Lambda$CDM: An Update”. New Astron. Rev., {\bf 95}:101659, (2022) \url{https://doi.org/10.1016/j.newar.2022.101659}.

\bibitem{Ferreira}
P.G, Ferreira, G. D. Starkman, ``Einstein’s Theory of Gravity and the Problem of Missing Mass", Science, {\bf 326}, 812 (2009) \url{https://ui.adsabs.harvard.edu/link_gateway/2009Sci...326..812F/doi:10.1126/science.1172245}

\bibitem{Robson}
B. A. Robson, ``The Matter-Antimatter Asymmetry Problem", JHEPGC, {\bf 4}, 1, (2018) \url{https://doi.org/10.4236/jhepgc.2018.41015}


\bibitem{Ancho}
L. A. Anchordoqui, I. Antoniadis, ``Large extra dimensions from higher-dimensional inflation", Phys. Rev. D, {\bf 109}, 103508 (2024) \url{https://doi.org/10.1103/PhysRevD.109.103508}

\bibitem{Teit}
C. Teitelboim, ``Quantum mechanics of the gravitational field", Phys. Rev. D ,{\bf 25}, 3159 (1982) \url{https://doi.org/10.1103/PhysRevD.25.3159}

\bibitem{Shankar}
S. Shankaranarayanan, J. P. Johnson, ``Modified theories of gravity: Why, how and what?", Gen. Relativ. Gravit., {\bf 54}, 44 (2022) \url{https://doi.org/10.1007/s10714-022-02927-2}

\bibitem{Quiros}
I. Quiros, ``Selected topics in scalar–tensor theories and beyond", Int. J. Mod. Phys. D, {\bf 28}, 07, 1930012 (2019) \url{https://doi.org/10.1142/S021827181930012X}


\bibitem{Guendelman0}
E. I. Guendelman, A. B. Kaganovich, ``Principle of nongravitating vacuum energy and some of its consequences", Phys. Rev. D {\bf 53}, 7020 (1996) \url{https://link.aps.org/doi/10.1103/PhysRevD.53.7020}


\bibitem{Guendelman1}
E. I. Guendelman and A. B. Kaganovich, ``Gravitational theory without the cosmological constant problem", Phys. Rev. D {\bf 55}, 5970, (1997) \url{https://doi.org/10.1103/PhysRevD.55.5970}

\bibitem{Guendelman2}
E.I. Guendelman, A.B. Kaganovich, ``Gravity, Cosmology and Particle Physics without the Cosmological Constant Problem" Mod. Phys. Lett. A, {\bf 13}, 19, 1583-1586 (1998) \url{https://doi.org/10.1142/S0217732398001662}

\bibitem{Guendelman3}
E.I. Guendelman, ``Scale Invariance, New Inflation and Decaying-terms", Mod. Phys. Lett. A, {\bf 14}, 16, 1043-1052 (1999) \url{https://doi.org/10.1142/S0217732399001103}

\bibitem{Guendelman4}
E.I. Guendelman, A.B. Kaganovich, ``Gravitational theory without the cosmological constant problem, symmetries of space-filling branes, and higher dimensions'', Phys. Rev. D, {\bf 56}, 6, 3548 (1997) \url{https://doi.org/10.1103/PhysRevD.56.3548}

\bibitem{Schutz}
B. F. Schutz, {\it A First Course in General Relativity (2nd ed.)}. Cambridge University Press, Cambridge (2009).

\bibitem{Weinberg}
S. Weinberg, {\it Gravitation and Cosmology: Principles and Applications of the General Theory of Relativity}, Wiley, (1972).

\bibitem{Blau}
M. Blau, ``Lecture Notes on General Relativity", (2022) \url{http://blau.itp.unibe.ch/GRLecturenotes.html}

\bibitem{Guendelman5}
E. I. Guendelman, R. Herrera, ``Curvaton reheating mechanism in a scale invariant two measures theory", Gen. Relativ. Gravit. {\bf 48}, 3 (2016) \url{https://doi.org/10.1007/s10714-015-1999-9}

\bibitem{Guendelman6}
E. I. Guendelman, R. Herrera, P. Labraña, ``Instant preheating in a scale invariant two measures theory", Phys. Rev. D, {\bf 103}, 123515 (2021) \url{https://doi.org/10.1103/PhysRevD.103.123515}

\bibitem{Guendelman7}
E. Guendelman, R. Herrera, ``Unification: Emergent universe followed by inflation and dark epochs from multi-field theory", Ann. Phys., {\bf 460}, 169566 (2024) \url{https://doi.org/10.1016/j.aop.2023.169566}










\bibitem{Chandrasekhar}
S. Chandrasekhar, The Mathematical Theory of Black Holes, Oxford University Press, New York (1983) \url{https://global.oup.com/academic/product/the-mathematical-theory-of-black-holes-9780198503705?cc=us&lang=en&}.

\bibitem{Weinberg2}
S. Weinberg, ``Cosmology", Indian Edition, Oxford University Press, New York (2008) \url{https://global.oup.com/academic/product/cosmology-9780198526827?cc=in&lang=en&}.

\bibitem{Hoyle}
F. Hoyle, ``A New Model for the Expanding Universe", Month. Not. Roy. Astron. Soc., {\bf 108}, 5, 372-382 (1948) \url{https://doi.org/10.1093/mnras/108.5.372}

\bibitem{Bondi}
 H. Bondi, T. Gold, ``The Steady-State Theory of the Expanding Universe", Month. Not. Roy. Astron. Soc., {\bf 108}, 3, 252-270 (1948) \url{https://doi.org/10.1093/mnras/108.3.252}
 
 \bibitem{Hoyle1}
F. Hoyle, G. Burbidge, J. Narlikar, ``A Quasi--Steady State Cosmological Model with Creation of Matter", Astrophys. J., {\bf 410}, 437 (1993) \url{https://ui.adsabs.harvard.edu/link_gateway/1993ApJ...410..437H/doi:10.1086/172761}.

\bibitem{Hoyle2}
F. Hoyle, G. Burbidge, J. Narlikar, ``Further astrophysical quantities expected in a quasi-steady state Universe", Astron. Astrophys., {\bf 289}, 729 (1994) \url{https://ui.adsabs.harvard.edu/abs/1994A%26A...289..729H/abstract}

\bibitem{Hoyle3}
 F. Hoyle, G. Burbidge, J. Narlikar, ``Astrophysical Deductions from the Quasi Steadystate Cosmology", Month. Not. Roy. Astron. Soc.,
 {\bf 267}, 1007 (1994) \url{https://ui.adsabs.harvard.edu/link_gateway/1994MNRAS.267.1007H/doi:10.1093/mnras/267.4.1007}.

 \bibitem{Hoyle4}
F. Hoyle, G. Burbidge, J. Narlikar, ``The basic theory underlying the quasi-steady-state cosmology", Proc. Roy. Soc. A, 448, {\bf 191} (1995) \url{https://doi.org/10.1098/rspa.1995.0012}



























\bibitem{Mukhanov}
V. Mukhanov, Physical Foundations of Cosmology, Cambridge Univer
sity Press, Cambridge (2005) \url{https://doi.org/10.1017/CBO9780511790553}.

\bibitem{Peebles}
P. J. E. Peebles, Principles of Physical Cosmology, Princeton University
 Press, New Jersey 1993, pp. 396 \url{https://press.princeton.edu/books/paperback/9780691209814/principles-of-physical-cosmology?srsltid=AfmBOophfTkIO_YHL2J6AidhhZFsSyjEKTymogiBhuBOj4BfWYVhaQFZ}.

\bibitem{Liddle}
A. R. Liddle, D. H. Lyth, Cosmological Inflation and Large-Scale Structure, Cambridge University Press, Cambridge (2000), pp. 49 \url{https://fma.if.usp.br/~mlima/teaching/PGF5292_2021/LiddleLyth_CILSS.pdf}.


\bibitem{Gray}
J. J. Gray and L. Tiling, ``Johann Heinrich Lambert, Mathematician and Scientist,” Historia Mathematica, {\bf 5}, 7, 13-14 (1978). \url{https://doi.org/10.1016/0315-0860(78)90133-7}

\bibitem{Corless}
R. M. Corless et. al., ``On the Lambert W Function”, Advances in Computational Mathematics, {\bf 5}, 1, 329-359 (1996) \url{http://dx.doi.org/10.1007/BF02124750}.

\bibitem{Majumder}
B. Majumder, S. Ray, and G. Manna, ``Evaporation of Dynamical Horizon with the Hawking Temperature in the K-essence Emergent Vaidya Spacetime",  Fortschr. Phys. {\bf 71}, 2300133, (2023) \url{https://doi.org/10.1002/prop.202300133}


\bibitem{Kaplinghat}
M. Kaplinghat et al., ``Observational constraints on power-law cosmologies", Phys. Rev. D {\bf 59}, 043514 (1999) \url{https://doi.org/10.1103/PhysRevD.59.043514}

\bibitem{Singh}
J.K. Singh, ``Power law cosmology in modified theory with thermodynamics analysis", Phys. Dark Univ., {\bf 46}, 101658 (2024) \url{https://doi.org/10.1016/j.dark.2024.101658}


\bibitem{Riess}
A. G. Riess et al., ``Observational Evidence from Supernovae for an Accelerating Universe and a Cosmological Constant", Astron. J., {\bf 116}, 1009 (1998) \url{https://iopscience.iop.org/article/10.1086/300499}

\bibitem{Perlmutter}
S. Perlmutter et al., ``Measurements of $\Omega$ and $\Lambda$ from 42 High-Redshift Supernovae", Astrophys. J. {\bf 517}, 565 (1999) \url{https://iopscience.iop.org/article/10.1086/307221}.

\bibitem{Komatsu}
E. Komatsu et al., ``Seven-year Wilkinson Microwave Anisotropy Probe (WMAP) Observations: Cosmological Interpretation", Astrophys. J. Suppl. {\bf 192}, 18 (2011) \url{https://iopscience.iop.org/article/10.1088/0067-0049/192/2/18}

\bibitem{Planck}
Planck Collab. (P. A. R. Ade et al.), ``Planck 2015 results XIV. Dark energy and modified gravity", Astron. Astrophys. {\bf 594}, A14 (2016) \url{	https://doi.org/10.1051/0004-6361/201525814}

\bibitem{Planck1}
Planck Collab. (N. Aghanim et al.), ``Planck 2018 results I. Overview and the cosmological legacy of Planck",  Astron. Astrophys. {\bf 641}, A1 (2020) \url{	https://doi.org/10.1051/0004-6361/201833880}

\bibitem{Planck2}
Planck Collab. (N. Aghanim et al.), ``Planck 2018 results VI. Cosmological parameters". Astron. Astrophys. {\bf 641}, A6 (2020) \url{	https://doi.org/10.1051/0004-6361/201833910}.

\bibitem{Frieman}
J. Frieman et al., ``Dark Energy and the Accelerating Universe",  Annual Review of Astronomy and Astrophysics,  {\bf 46}, 385-432 (2008) \url{https://doi.org/10.1146/annurev.astro.46.060407.145243}  

\bibitem{Nemiroff}
R. J. Nemiroff, B. Patla,``Adventures in Friedmann cosmology: A detailed expansion of the cosmological Friedmann equations" Am. J. Phys. {\bf 76}, 265–276 (2008) \url{https://doi.org/10.1119/1.2830536}

\bibitem{Smeulders}
P. Smeulders, ``Why the Expansion of the Universe Appears to Accelerate", Journal of Modern Physics, {\bf 4}, 6 (2013) \url{http://dx.doi.org/10.4236/jmp.2013.46107}

\bibitem{Kim}
H. Kim, ``Brans-Dicke theory as a unified model for dark matter-dark energy", Mon. Not. Royal Astron. Soc., {\bf 364}, 3, 813–822 (2005) \url{https://doi.org/10.1111/j.1365-2966.2005.09593.x}

\bibitem{Albareti2012}
F.D. Albareti et al., ``Focusing of geodesic congruences in an accelerated expanding Universe", JCAP, {\bf 12}, 020 (2012) \url{https://iopscience.iop.org/article/10.1088/1475-7516/2012/12/020}

\bibitem{Das1}
S. Das, ``Quantum Raychaudhuri equation", Phys. Rev. D {\bf 89}, 084068 (2014) \url{https://doi.org/10.1103/PhysRevD.89.084068}

\bibitem{Senovilla}
Jose M.M. Senovilla, ``New class of inhomogeneous cosmological perfect-fluid solutions without big-bang singularity", Phys. Rev. Lett. {\bf 64} 2219-2221, (1990) \url{https://doi.org/10.1103/PhysRevLett.64.2219}.

\bibitem{Senovilla1}
Jose M.M. Senovilla, ``Singularity Theorems and Their Consequences", Gen. Rel. Grav., {\bf 30}, 701-848 (1998) \url{https://doi.org/10.1023/A:1018801101244}

\bibitem{Maddox}
John Maddox, ``Another gravitational solution found", Nature {\bf 345}, 201, (1990) \url{https://doi.org/10.1038/345201a0}

\bibitem{Bhattacharyya}
I. Bhattacharyya and S. Ray, ``Accelerated motion in general relativity: fate of the singularity", Eur. Phys. J. C {\bf 82}, 953 (2022) \url{https://doi.org/10.1140/epjc/s10052-022-10876-y}

\bibitem{Choudhury2}
S. G. Choudhury et al., ``Self-similar collapse and the Raychaudhuri equation", Eur. Phys. J. C {\bf 79}, 1027 (2019) \url{https://doi.org/10.1140/epjc/s10052-019-7559-9}.






































\end{thebibliography}
\end{document}